\documentclass[iop,apj,numberedappendix,twocolumn, tighten]{aastex62}

\usepackage{natbib}
\usepackage{graphicx}
\usepackage{amsmath,amsfonts}
\usepackage{multirow}
\usepackage{xspace}
\usepackage[normalem]{ulem}
\usepackage{booktabs}
\usepackage{colortbl}
\usepackage{hyperref}
\usepackage[nameinlink]{cleveref}
\usepackage[encapsulated]{CJK}
\usepackage{ucs}
\usepackage{float}
\usepackage[utf8x]{inputenc}

\usepackage{eht}
\usepackage{newtxtext} 
\usepackage{mathrsfs}
\usepackage{lineno}
\usepackage{xcolor}

\usepackage{rotating}
\usepackage{array}
\usepackage{arydshln}
\definecolor{fed_blue}{HTML}{07004D}
\definecolor{steel_blue}{HTML}{2D82B7}
\definecolor{aqua_marine}{HTML}{42E2B8}
\definecolor{dutch_white}{HTML}{F3DFBF}
\definecolor{light_coral}{HTML}{EB8A90}

\hypersetup{
    colorlinks = true,
    linkcolor = light_coral,
    urlcolor  = steel_blue,
    citecolor = steel_blue,
    anchorcolor = black
}

\graphicspath{{./figures}}

\newcommand{\muas}{$\mu$as\xspace}

\newcommand{\kerrbam}{\texttt{KerrBAM}\xspace}

\renewcommand{\vec}[1]{\mathbf{#1}}

\def\sgra{Sgr~A$^{\ast}$\xspace}
\def\virgoa{M87$^{\ast}$\xspace}
\def\mring{$m$-ring\xspace}
\def\lsim{\mathrel{\raise.3ex\h box{$<$\kern-.75em\lower1ex\hbox{$\sim$}}}}
\def\gsim{\mathrel{\raise.3ex\hbox{$>$\kern-.75em\lower1ex\hbox{$\sim$}}}}
\def\gtwid{\mathrel{\raise.3ex\hbox{$>$\kern-.75em\lower1ex\hbox{$\sim$}}}}
\def\proptwid{\mathrel{\raise.3ex\hbox{$\propto$\kern-.75em\lower1ex\hbox{$\sim$}}}}

\defcitealias{TMS}{TMS}
\setcitestyle{notesep={; }}

\begin{document}

%% TITLE %%
\title{\textbf{Machine- and deep-learning-driven angular momentum inference from BHEX observations of the $n=1$ photon ring}} 
% \title{\textbf{Angular momentum inference from BHEX images: a framework for $n=1$ photon subring analysis using machine and deep learning approaches}} 
\shorttitle{$n=1$ Photon Ring Study}

%% AUTHORS %%
\author[0000-0003-4914-5625]{Joseph Farah}
\affiliation{Las Cumbres Observatory, 6740 Cortona Drive, Suite 102, Goleta, 
CA 93117-5575, USA}
\affiliation{Department of Physics, University of California, Santa Barbara, 
CA 93106-9530, USA}

\author[0000-0002-2685-2434]{Jordy Davelaar}
\affiliation{Department of Astrophysical Sciences, Peyton Hall, Princeton University, Princeton, NJ 08544, USA}
\affiliation{NASA Hubble Fellowship Program, Einstein Fellow}
\affiliation{Center for Computational Astrophysics, Flatiron Institute, 162 Fifth Avenue, New York, NY 10010, USA}
\affiliation{Department of Astronomy and Columbia Astrophysics Laboratory, Columbia University, 550 W 120th St, New York, NY 10027, USA}
\author[0000-0002-7179-3816]{Daniel Palumbo}
\affiliation{Center for Astrophysics | Harvard \& Smithsonian, 60 Garden Street, Cambridge, MA 02138,
USA}
\affiliation{Black Hole Initiative at Harvard University, 20 Garden Street, Cambridge, MA 02138, USA}
\author[0000-0002-4120-3029]{Michael Johnson}
\affiliation{Center for Astrophysics | Harvard \& Smithsonian, 60 Garden Street, Cambridge, MA 02138,
USA}
\affiliation{Black Hole Initiative at Harvard University, 20 Garden Street, Cambridge, MA 02138, USA}
\author[0009-0005-8120-8499]{Jonathan Delgado}
\affiliation{Department of Mathematics, University of California, Irvine, CA 92697, USA}
% \author{others}

\shortauthors{Farah et al.}

%% CORRESPONDENCE %%
\correspondingauthor{$^\dag$Joseph R. Farah}
\email{josephfarah@ucsb.edu}

%% ABSTRACT %%
\begin{abstract}
The $n=1$ photon ring is an important probe of black hole (BH) properties and will be resolved by the Black Hole Explorer (BHEX) for the first time. However, extraction of black hole parameters from observations of the $n=1$ subring is not trivial. Developing this capability can be achieved by building a sample of $n=1$ subring simulations, as well as by performing feature extraction on this high-volume sample to track changes in the geometry, which presents significant computational challenges. Here, we present a framework for the study of $n=1$ photon ring behavior and BH property measurement from BHEX images. We use \texttt{KerrBAM} to generate a grid of $\gtrsim10^6$ images of $n=1$ photon rings spanning the entire space of Kerr BH spins and inclinations. Intensity profiles are extracted from images using a novel feature extraction method developed specifically for BHEX. This novel method is highly optimized and outperforms existing EHT methods by a factor of ${\sim}3000$. Additionally, we propose a novel, minimal set of geometric measurables for characterizing the behavior of the $n=1$ subring geometry. We apply these measurables to our simulation grid and test spin recovery on simulated images using: (i) gradient boosting, a machine learning algorithm; and (ii) an extension of Deep Horizon, a deep learning framework. We find $\gtrsim90$\% correct recovery of BH properties using the machine/deep learning approaches, and characterize the space of resolution-dependent geometric degeneracies. Finally, we test both approaches on GRMHD simulations of black hole accretion flows, and report accurate recovery of spin at the expected inclination of \virgoa.
\end{abstract}

%% KEYWORDS %%

%% SECTIONS %%
\section{Introduction}
\label{sec:introduction}

% First image of a black hole and upcoming BHEX
In April 2019, the Event Horizon Telescope (EHT) Collaboration published the first image of a black hole \citep{EHT1,EHT4}, representing a significant step forward in resolving and imaging astrophysical phenomena via very-long-baseline interferometry (VLBI). The unprecedented resolution of the image ($\approx 20$ \muas) uniquely probed event-horizon scale phenomena and directly facilitated the strongest tests of general relativity to-date \citep[e.g.,][etc.]{EHT6,EHT7,Bambi2019,Psaltis2018}. However, while this remarkable resolution was sufficient to probe the shadows of the \virgoa and \sgra black holes \citep{SgrAPaperI,SgrAPaperIII}, it was still inadequate to probe the more complex features of event-horizon scale emission predicted by theory, such as the photon ring. Recently, a space-baseline extension to the EHT (the Black Hole Explorer; BHEX) has been proposed to overcome these limitations and help resolve open questions about general relativity and other fundamental physics \citep{BHEXVision,BHEXJapan}.

% The advanced resolution of BHEX will enable us to see the n=1 photon ring, which proxies directly to interesting black hole properties
In particular, the higher resolution of BHEX will enable direct observation and study of a black hole's photon ring \citep{BHEXPRS}, a shell of unstable photon orbits which, in aggregate, produce a sharp emission feature on the gravitationally lensed image of the black hole seen by a distant observer. The photons completing $n$ half-orbits around the black hole produce $n$ distinct and nested subrings \citep{Johnson2020}. Of these subrings, the $n=0$ direct image is primarily determined by the fluid flow and only partially determined by the spacetime. By contrast, the morphology of the $n\geq1$ subrings (the ``photon rings'') is only dictated by unstable photon orbits and are therefore mostly dependent on the properties of the black hole spacetime (e.g., spin, inclination relative to the observer, etc.) \citep[e.g.,][]{BroderickPRS}. The resolution of BHEX is expected to be sufficient to probe at least the $n=1$ subring, thus enabling direct and precise measurements of black hole properties, particularly spin \citep{Johnson2020}.

% The behavior of the n=1 photon ring is poorly understood, beyond characterizations of size. (issues with performance/feature extraction)
The behavior of the $n=1$ photon ring as modified by spin is poorly understood, presenting a potential hurdle for precision measurements via BHEX observations \citep[see e.g.,][for a review]{beginnersguide}. Unlike the crticial curve, which shows substantial deviation from circularity at edge-on inclination and spins approaching unity \citep[e.g.,][etc]{Bardeen1973,vries_2003,Cunha_2018,Farah2020,Gralla2020}, the equivalent modification to the $n=1$ subring is more complex and more subtle. Previous studies have characterized the overall size and ellipticity \citep{BroderickPRS,Paugnat2022,Gralla2020PRSShape}, which varies with the angular momentum of the black hole. The azimuthal brightness modulation of the $n=1$ subring, which can be modeled by a shape known as an \mring \citep{Johnson2020}, also varies with spin. The precise relationship between these quantities and the properties of the black hole is still unknown; however, characterizing these relationships is essential for measurements with detections of the first subring.

The limited ability to observe such objects in nature results in a heavy reliance on simulations \citep[e.g.,][]{EHT5,EHT6,SgrAPaperV,SgrAPaperVI}, which are computationally challenging to perform with existing tools. In particular, in order to study the $n=1$ subring properties, we rely on computational methods such as has general relativistic magnetohydrodynamic (GRMHD) simulations, which are computationally expensive to perform for the full parameter space in e.g. spin. In addition, image feature extraction tools developed for EHT applications involving a small number of images are robust but have generally poor scalability, requiring weeks or even months to analyze the large ($\gtrsim10^5$) number of images needed for machine-learning-enabled analyses.

% In this paper, we propose
In this paper, we present a novel framework to enable large-scale studies of the $n=1$ photon subring. We exploit the fast semi-analytic simulation software \kerrbam \citep{BAM} to simulate the first subring and develop bespoke tools to perform analyses on feasible timescales. In \autoref{sec:data}, we describe the efficient generation of a suite of $n=1$ photon rings spanning all possible spins and inclinations using \kerrbam, a recently-developed semi-analytic accretion modeling code. In \autoref{sec:rbpe}, we introduce and evaluate ``recursive brightest-point extraction'', a novel feature extraction approach designed specifically for BHEX and built to be thousands of times more computationally efficient than existing tools. In \autoref{sec:ellipse}, we introduce a set of non-redundant geometric observables which can characterize the shape and emission characteristics of the various subrings in our simulation grid. We use these geometric observables to provide the first numerical analysis of the behavior of the $n=1$ subring across the Kerr parameter space of spins and inclinations. In \autoref{sec:fitting} we demonstrate the ability of these observables as an unambiguous measurement tool by training machine- and deep-learning algorithms to estimate black hole properties from both numerical observables as well as full images of subrings simulated with \kerrbam. Finally, in \autoref{sec:bhex_app} we directly apply our methods to GRMHD-simulated black holes. We summarize our conclusions and outlook for BHEX measurements in \autoref{sec:conclusions}.

% \clearpage
\section{Simulation generation via \kerrbam}
\label{sec:data}

% FIGURES: (1) grid of simulations, (2) n0, n1, and n1+n0
% TABLES: (1) table of KerrBAM parameters

% summarize need for simulation library
The inaccessibility of the $n=1$ photon ring given current observational limitations results in our understanding being theoretical and largely simulation-based \citep[e.g.,][]{EHT5,EHT6,SgrAPaperV,SgrAPaperVI}. Machine learning algorithms in particular benefit from large datasets to improve performance, particularly when generalizing training features to real-world behavior or attempting to recover numerical quantities in a continuous fashion \citep{ElementsStatistics}. 
Full GRMHD simulations, while frequently used and closer to reality than semi-analytic simulations, are also computationally expensive to generate, presenting an early bottleneck in attempts to characterize subtle feature behavior. Fortunately, recent years has seen the development of new models and techniques for analytic or semi-analytic computation and/or approximation of black hole accretion flows, which have come with a substantial increase in performance \citep{ART2,BAM,ART1}. 

% summarize KerrBAM
We generated a simulated dataset of $10^6$ images to use for training. This is a substantial ($10$x) increase in dataset size as compared to previous similar analyses \citep[e.g., Deep Horizon:][]{Jordy}. The dataset size is motivated by the subtlety of the subring behavior we hope to capture; the spin and inclination parameter space has steep gradients (particularly towards maximal spin or edge-on inclination) which require dense sampling to characterize. The simulated images are generated with the \kerrbam simulator, a semi-analytic ray tracing engine capable of producing images of all $n\geq0$ subrings \citep{BAM}. We fixed all accretion flow parameters to standard values following \cite{BAM}. For the parameters controlling the black hole spacetime, we fixed all parameters to values consistent with known values for \virgoa, with the exception of angular momentum (spin) $a$, inclination $\theta_0$, and emission radius. The fixed values are summarized in \autoref{tab:bh_params}. Spin was varied between almost minimal ($|a|=0.04$, in units of $Jc/M^2$, where $J$ is the angular momentum of the black hole) and almost maximal ($|a|=0.94$) for both negative (spin axis oriented away from observer) and positive (spin axis oriented towards observer) spin. Inclination was varied between face-on ($\theta_0=0^\circ$, parallel to the spin axis) and near-edge-on ($\theta_0=85^\circ$, almost perpendicular to the spin axis). The emission radius was varied between $3$ $M$ (where $M$ is equivalent to the Schwarzschild radius of the black hole via $M\sim r_s c^2/2G$) and $7$ $M$. Images were generated with adaptive ray tracing enabled and resized to a field-of-view (FOV) of $80$ \muas and resolution 128x128 pixels using a linear spline interpolation. 

\begin{table}
    \begin{centering}
    \begin{tabular}{ccccc}
    % \bottomrule
    \bottomrule
    \textbf{Parameter} & FOV$^a$ & $M^b$ & $D^c$ & PA$^d$ \\ \cdashline{1-5}
    \textbf{Value} & 80 \muas & $6.5\times10^9 M_\odot$ & $5.1\times10^{23}$ m & $0^\circ$ \\ 
    \toprule
    \end{tabular}
    \end{centering}
    $^a$The field-of-view of the output image. \\
    $^b$The mass of the simulated black hole, set to the \virgoa mass. \\
    $^c$The distance between the observer and the simulated black hole, set to the \virgoa distance. \\
    $^d$The position angle of the spin axis projected in the observer plane, corresponding to a rotation of the image.\\
    \caption{\kerrbam simulation parameters controlling the black hole spacetime. We cite \cite{M87PaperIV} where needed for relevant values.}
    \label{tab:bh_params}

\end{table}

\begin{figure*}[]
    \centering
    \includegraphics[]{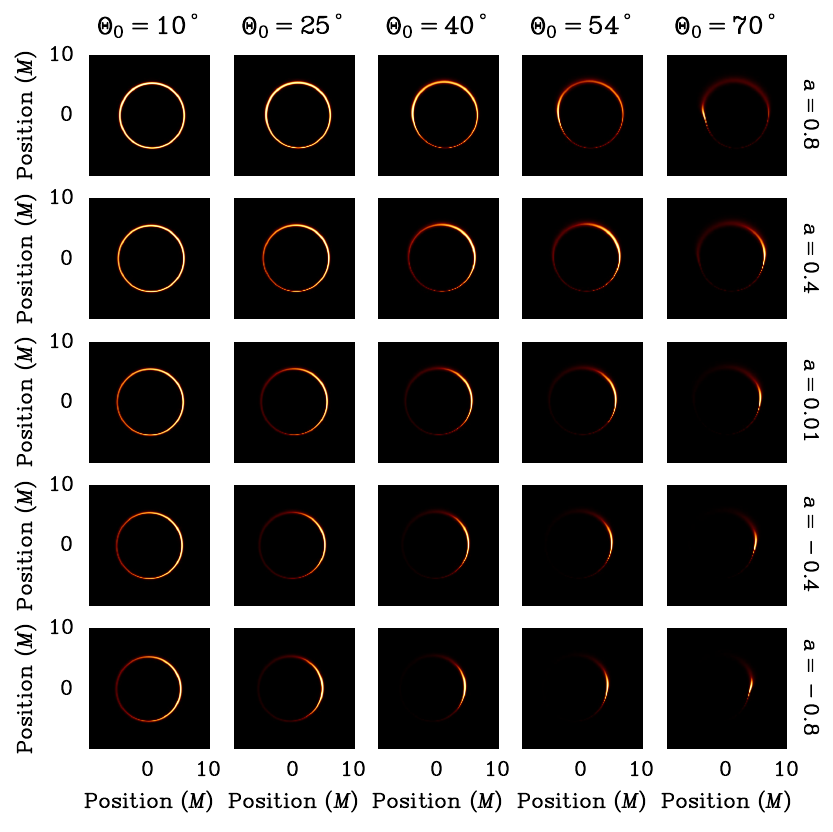}
    \caption{Sampled $n=1$ photon rings generated with \kerrbam. Our dataset consists of $\gtrsim10^6$ images spanning almost the entire Kerr parameter space.}
    \label{fig:dataset_demo}
\end{figure*}

% describe the images 
A sample set of images from the dataset are shown in \autoref{fig:dataset_demo}. These images show the varying behavior of the $n=1$ photon ring the $n=1$ photon ring for varying black hole parameters. At edge-on inclination, the brightness and circular asymmetry becomes more pronounced at higher spin. However, at face-on inclinations, this behavior is signifanctly less prominent. The variation in spin at these low inclinations is more apparent as a change in the brightness modulation around the ring.
%In \jrf{FIGURE}, we show how these $n\geq1$ images aggregate with the more accessible $n=0$ emission to produce a the full lensed projection of the black hole and surrounding accretion flow.

% \clearpage
\section{Recursive brightest-point extraction}
\label{sec:rbpe}

% FIGURES: (1) flowchart of recursive algorithm, (2) visualization of iterations, (3) performance comparison vs. REX 

% summary of goal with extraction
For this analysis, we seek to quantify the emission profile and shape of the $n=1$ subring, in order to assess how it varies with spin. Image-domain, ring-based feature extraction has been developed for previous VLBI imaging analyses and has enabled, among other things, size and mass measurements of the \virgoa black hole based on images of the shadow \citep{Johannsen2016,eht_imaging}. Once a series of points corresponding to the brightness profile has been measured, Bayesian inference methods can be used to quantify the geometry of the ring shape. By quantifying the modification of the shape with spin and inclination, we can make assessments about the feasibility of using the subring shape to recover black hole parameters in data taken by BHEX.

% explain ray-based extraction (REX)
The most common method for extracting a ring-like brightness profile involves identifying peak points of flux along rays emanating in all directions from an assumed center. This method was used to make measurements of the black hole shadows of \virgoa and \sgra \citep{M87PaperIV,SgrAPaperIII}. In some implementations, the assumed center is fixed and is not recalculated; however, in others, the assumed center is recalculated afterwards to minimize the average peak distance from the center. We use the ray-based feature extraction implementation in \texttt{eht-imaging}, specifically the \texttt{REX} module, to test the performance and efficiency of ray-based feature extraction \citep{eht_imaging}. We find in practice that ray-based feature extraction has several limitations that present potential obstacles to our analysis. First, due to the need to scan along dozens of rays and perform several objective function minimizations, this method of feature extraction is slow, with peak flux extractions ranging from $\approx1\mathrm{-}10$ seconds. This time range would result in a required time budget of $\approx10\mathrm{-}100$ days to fully analyze our $10^6$ simulated images. Additionally, the method is somewhat dependent on the assumption of circularity, which is not consistently met by images of the $n=1$ subring. Particularly at high spin, the shape may be significantly asymmetrical in both brightness distribution and ellipticity, resulting in incorrect or incomplete extractions.

\begin{figure*}
    \centering
    \includegraphics[scale=0.19]{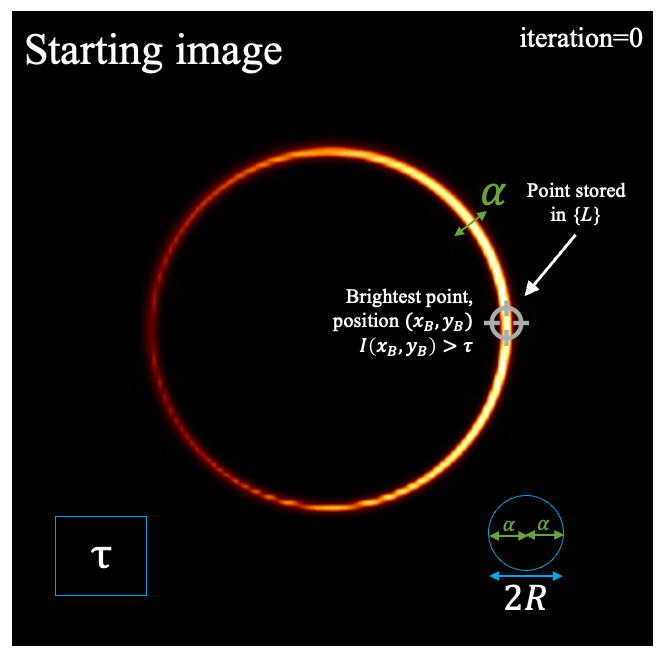}
    \includegraphics[scale=0.19]{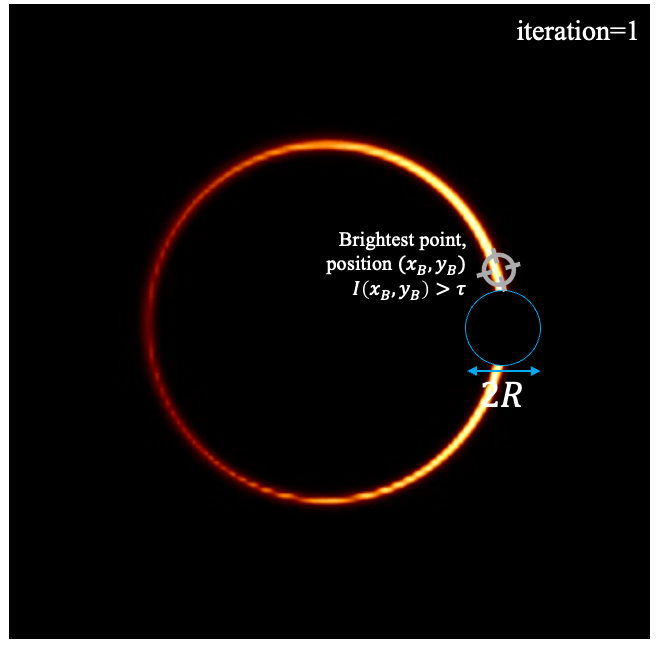}
    \includegraphics[scale=0.19]{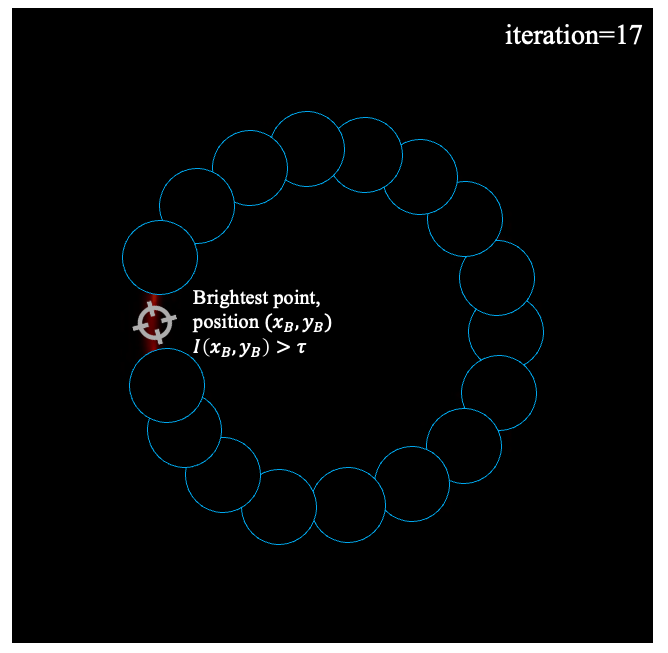}
    \includegraphics[scale=0.19]{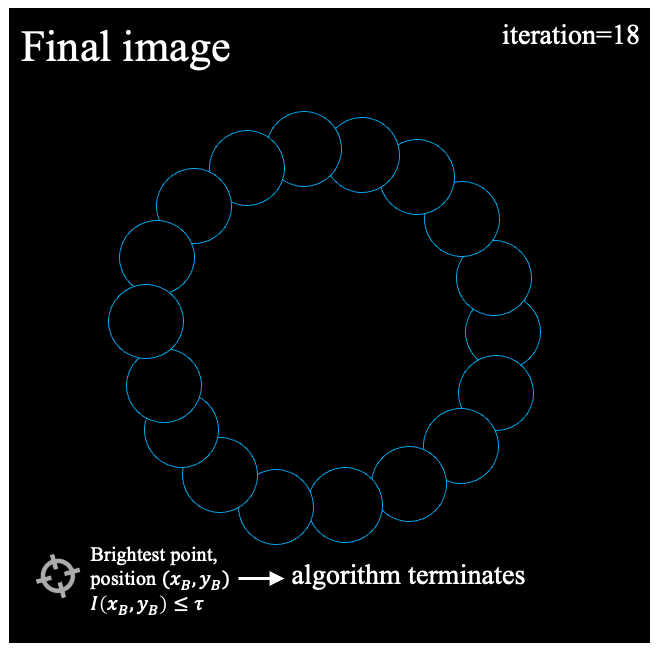}
    \caption{Visualization of the recursive brightest-point (RBP) extraction algorithm. The starting image configuration (far left) is used to estimate acceptable values for the algorithm parameters. The masking radius $\mathcal{R}$ is set to some value greater than the average radial emission profile width $\alpha$. The noise stopping criterion $\tau$ is estimated based on a $5\sigma$ upper limit on the background flux distribution, sampled from within the blue square region. For the zeroeth iteration, the brightest point, at position $(x_B, y_B)$ is found and added to $\{\mathcal{L}\}$. A region of radius $\mathcal{R}$ is masked out centered on $(x_B, y_B)$, and the masked image is recursively passed to the algorithm. For the next iteration, the next brightest point is found and added to $\{\mathcal{L}\}$, and the masking is repeated. This process is repeated for 17 iterations (middle right), during which the final point on the ring is masked out. On the final iteration (far right), all meaningful emission from the $n=1$ photon ring has been masked out and the only flux left is noise. The brightest point in the image now has an intensity less than our stopping criterion $\tau$, so the algorithm terminates. }
    \label{rbp:demo}
\end{figure*}

% introduce recursive brightest-point extraction
To address these limitations, we have developed a novel and bespoke method of feature extraction designed for maximum robustness and performance specifically on BHEX subring images. The new algorithm distills the process of extracting peak fluxes into two vectorized steps which are then applied recursively. Given a BHEX image $I(x,y)$ of a photon subring with emission following some arbitrary parameterized curve $C(t)$, a characteristic angular radius $\mathcal{R}$, an intensity threshold $\tau$, and an initially empty list of emission profile points $\{\mathcal{L}\}$, the algorithm is as follows:
\begin{enumerate}   
    \item Identify the brightest point $(x_B, y_B)$ in the image based on intensity $I(x,y)$. 
    \begin{enumerate}
        \item \texttt{IF} $I(x_B,y_B) \leq \tau$: this is considered noise. \texttt{QUIT} and \texttt{RETURN} the list of points $\{\mathcal{L}\}$ corresponding to the emission profile.
        \item \texttt{IF} $I > \tau$: The coordinate $(x_B, y_B)$ is considered a peak flux location; \texttt{ADD} to the list $\{\mathcal{L}\}$ corresponding to the emission profile of the image.
    \end{enumerate}
    \item Mask out a circular region of radius $\mathcal{R}$ and center $(x_B,y_B)$ in the image.
    \item \texttt{RETURN} the masked image, the updated list of points $\{\mathcal{L}\}$, the radius $\mathcal{R}$, and the threshold $\tau$ to the function recursively.
\end{enumerate} 
A visualization of this algorithm is shown in \autoref{rbp:demo}. The masking radius $\mathcal{R}$ is chosen based on the estimated width $\alpha$ of the emission profile; a choice of $\mathcal{R} < \alpha$ may result in an incorrect recovery. When $\mathcal{R} \geq \alpha$, $\mathcal{R}$ controls the density of the points recovered along the ring. For an emission profile total integrated length $\ell = \int_{C} dt$, the number of points $N$ produced with a chosen $\mathcal{R} \geq \alpha$ will be, on average, $N\sim\ell/\mathcal{R}$. The choice of $\tau$ is more complex but simultaneously less impactful; an optimal choice would be slightly lower than the dimmest point on the desired emission curve, but this is difficult to ascertain prior to calculating the emission profile. Another natural choice is simply an estimate of the average background intensity, which is straightforward to calculate quickly. As the emission profile is repeatedly masked out by step (2) in the algorithm, leaving only the background, the average background intensity served as an effective stopping criterion in our testing.

% compare to REX
We implement our novel recursive brightest-point (RBP) extraction algorithm in \texttt{python} using vectorized tools in the \texttt{scipy} scientific computing package and compare performance to \texttt{REX}. The short, recursive nature of the RBP algorithm, combined with the full vectorization, results in a speed increase of $\approx1\textrm{-}3\times10^3$x over ray-based feature extraction. On average, the RBP algorithm takes $\sim10^{-3}$ seconds per extraction, compared to $\approx1$ second for \texttt{REX}. The spread in this extraction time for RBP is also small, as the performance of RBP is completely unaffected by the geometry of the image ($\sim1\times10^{-4}$). By contrast, \texttt{REX} tends to perform worse (both in terms of time complexity and extraction success) on images with higher spins and inclinations, as the image becomes less circular and the azimuthal modulation of the brightness becomes more asymmetric. 

However, RBP is strongly sensitive to the assumption that the brightest points in the image lie on the $n=1$ subring. Other emission present--for example, in an aggregate image containing all subrings--will cause incorrect extractions using RBP. By contrast, \texttt{REX} scans for the brightest point along a particular ray, which the $n=1$ subring emission will frequently satisfy. In practice, RBP is useful for developing large training sets on clean images where speed is a priority, while \texttt{REX} is better suited for more complex individual cases (e.g., real image reconstructions), where the time budget is greater but greater robustness is desired.

\section{Geometric measurables}
\label{sec:ellipse}

% FIGURES: (1) master figure showing how main measurables are derived, (2) demo on low spin vs. high spin

% need to quantify shape and considerations and limitations for measurables
In order to quantify the transformation of the $n=1$ subring shape with spin and inclination, we next choose and calculate geometry-characterizing statistics (``geometric measurables'') on the extracted peak flux locations. For the purposes of training a machine learning algorithm, we seek to identify the smallest set of geometric measurables that can successfully distinguish between subrings at different spins. Ideally, these measurables will be computable in $\mathcal{O}(n)$ (linear) or $\mathcal{O}(n \log n)$ (logarithmic) time complexity, to minimize the time required to generate training data. Additionally, these measurables cannot be constructed in reference to any absolute coordinate system, as an absolute phase reference will likely be unavailable to BHEX. 

% propose set of measurables and verify performance, demonstration
The shape of the $n=1$ subring is well-approximated by an ellipse for a large majority of the $a\textrm{-}\theta_0$ parameter space. An ellipse fit encodes useful information, but the iterative nature of any objective function minimization routine makes a fit sub-optimal for performance. As an alternative, we use the \textit{smallest enclosing ellipse} (SEE), the ellipse of smallest area which encloses all recovered points. The SEE can be calculated recursively in $\mathcal{O}(n)$ time \citep{SEE_ref}.
Additionally, the SEE has a much higher degree of variation with spin than a corresponding ellipse fit, making it a more effective diagnostic. This higher degree of variation is caused by the SEE encoding information about the brightness modulation around the ring, but also makes the SEE diagnostic more sensitive to the dynamic range of the interferometer. To construct our geometric measurables, we define the following quantities: $r_a$, the semimajor axis of the SEE; $r_b$, the semiminor axis of the SEE; $\varphi_{\mathrm{S}}$, the principal angle of the SEE; and $\vec{C}_{\mathrm{S}}=(x_{ab}, y_{ab})$, the centroid of the SEE. We additionally construct the center-of-mass $\vec{C}_m$ of the peak flux locations $\{\mathcal{L}\}_{i=0}^N$ with
\begin{equation}
    \vec{C}_m = 
    \begin{bmatrix}
        C_x \\
        C_y 
    \end{bmatrix}  
    = 
    \begin{bmatrix}
        \frac{1}{N}\sum_i L_i^x \\
        \frac{1}{N}\sum_i L_i^y
    \end{bmatrix} ,  
\end{equation} 
where $L_i^x$ denotes the $x$ component of a point identified by the RBP algorithm. We propose the following geometric measurables to comply with the constraints discussed above:
\begin{enumerate}
     % \item $r_a$: the semimajor axis of the SEE,
     % \item $r_b$: semiminor axis of the SEE,
     \item $\varepsilon$: an eccentricity parameter, constructed via $$\varepsilon \equiv \frac{1}{r_a}\sqrt{r_a^2 - r_b^2},$$
     \item $|\vec{C}_{\mathrm{m}} - \vec{C}_{\mathrm{S}}|/r_a$: Magnitude of vector between SEE centroid and center-of-mass of points, and
     \item $\varphi_{\mathrm{S}}-\arg(\vec{C}_{\mathrm{m}} - \vec{C}_{\mathrm{S}})$: Angle of SEE relative to vector between centroid and center-of-mass of points. 
\end{enumerate} 
These geometric measurables can be computed with linear time complexity, and are generally non-redundant, making them ideal for both generating and fitting our $10^6$ element training set. Critically, these measurables are also scaled to a proxy of the overall size of the subring image ($r_a$), making them insensitive to the $M/D$ ratio or the emission radius of the target black hole. In addition to these measurables, we also fit an ellipse with semimajor axis $a_e$, semiminor axis $b_e$, and principal angle $\varphi_e$ to the peak flux locations $\{\mathcal{L}\}$ for the purpose of characterizing the $n=1$ subring shape independent of the brightness modulation. These quantities ($a_e, b_e, \varphi_e$) are not used for training, and are computed for a smaller ($\sim3\times10^4$) grid of images spanning the $a\textrm{-}\theta_0$ parameter space for comparison to the SEE measurables.

\subsection{Numerical variation with black hole properties}
\label{sec:num_var}

% FIGURES: (1) Plots of variation of numerical quantities with spin and inclination

\begin{figure*}[]
    \centering
    \includegraphics[scale=0.5]{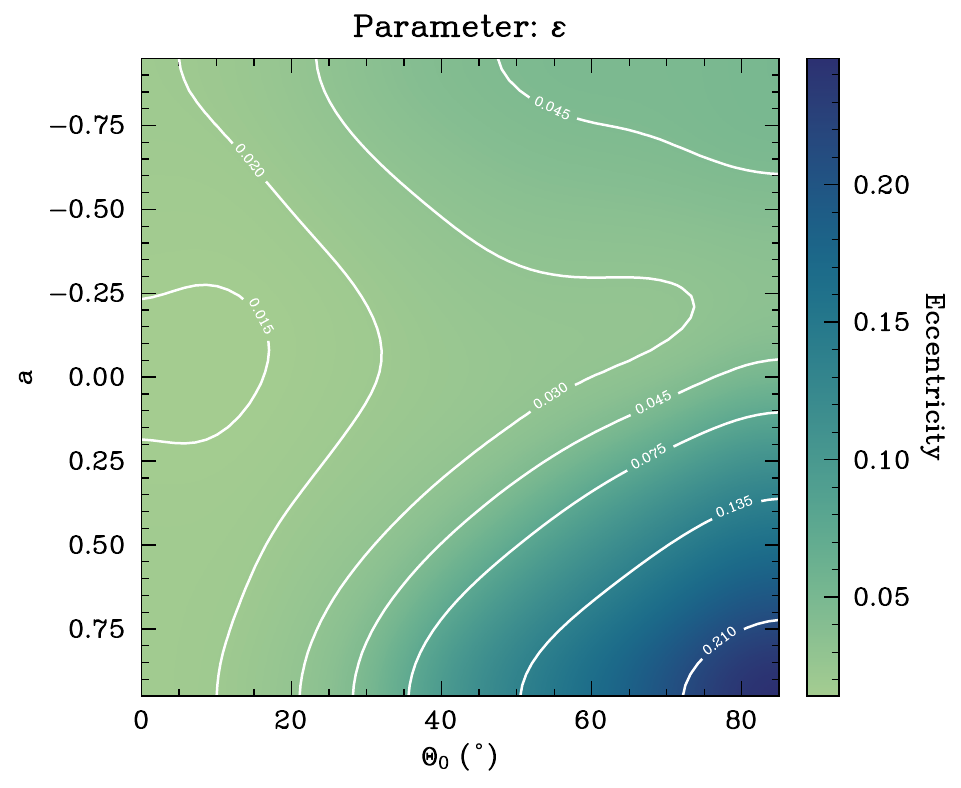}
    \includegraphics[scale=0.5]{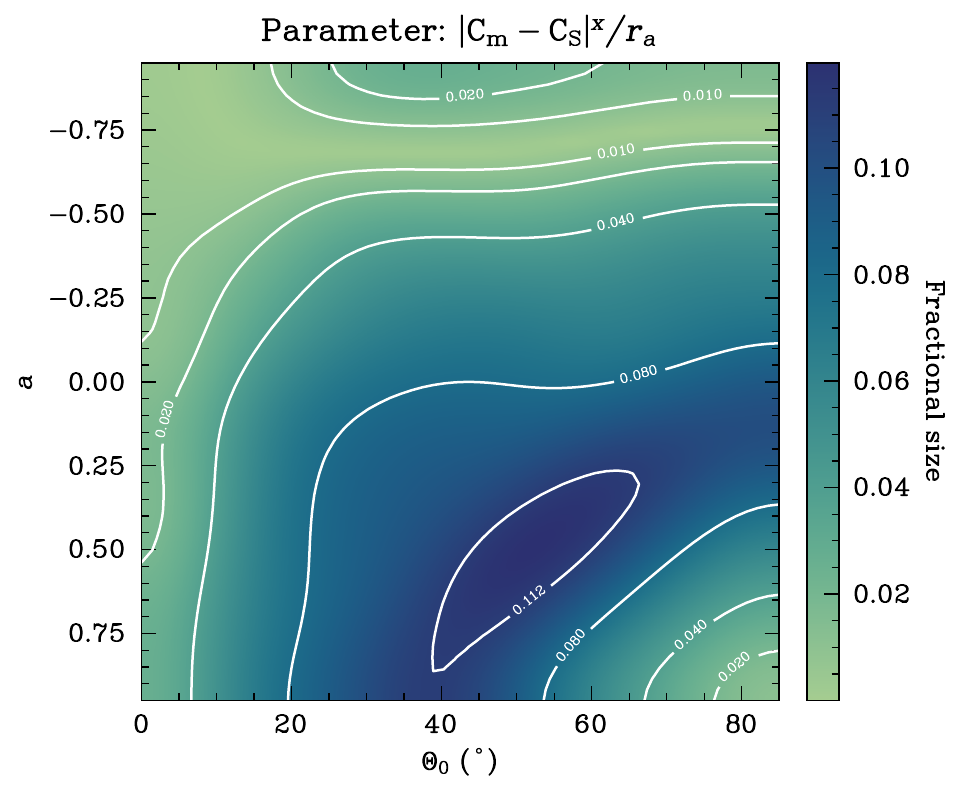}
    \includegraphics[scale=0.5]{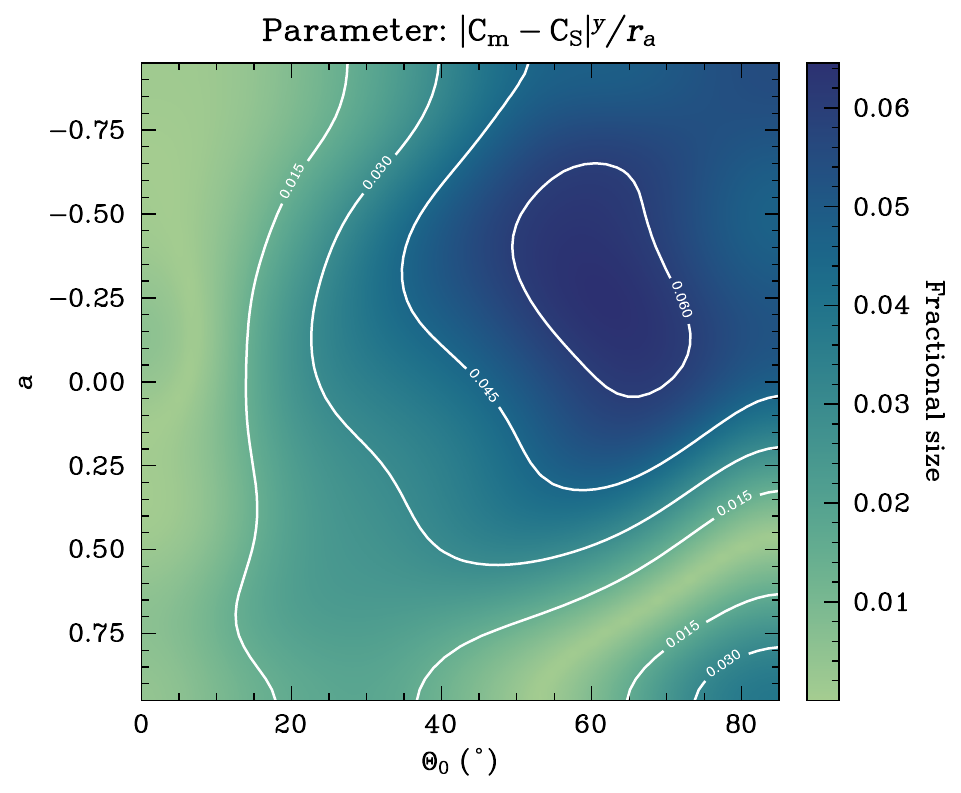}
    \includegraphics[scale=0.5]{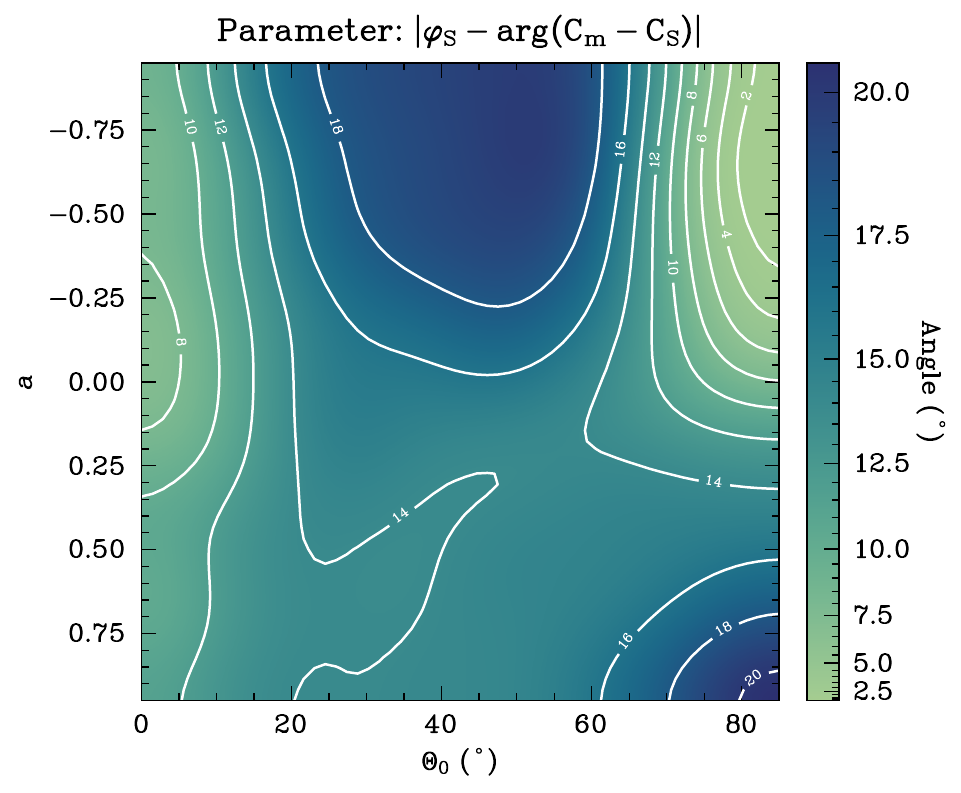}
    \caption{Heatmaps of the SEE quantities described in \autoref{sec:ellipse}. }
    \label{fig:heatmap_training_params}
\end{figure*}

% describe heatmap plots for training quantities
We compute each of the quantities described above for the images in our dataset and plot a heatmap of each result in \autoref{fig:heatmap_training_params}.% We additionally plot the best-fit ellipse parameters for each simulation in \autoref{fig:heatmap_training_params_ellipse}.
Broadly, all geometric measurables show relatively smooth and simple behavior modification with spin and inclination. At inclinations below $\theta_0\approx40^\circ$, much of the geometric behavior of the rings is highly degenerate, indicating likely significant challenges distinguishing spin at low inclination (such as that of \virgoa). Most parameters tend to an extreme value in the limit of high spin and edge-on inclination, except for the vertical component of $|\vec{C}_{\mathrm{m}} - \vec{C}_{\mathrm{S}}|$, which becomes most extreme towards low spin and high inclination. After the $\theta_0\gtrsim 40^\circ$ boundary, variation in the parameters is fairly linear up to maximum spin and edge-on inclination. In general, all parameters varied substantially more for negative spins than for positive spins. 

% \begin{figure*}[]
%     \centering
%     \includegraphics[scale=0.5]{figures/num_analysis/ellipse/numerical_analysis_param=a.pdf}
%     \includegraphics[scale=0.5]{figures/num_analysis/ellipse/numerical_analysis_param=b.pdf}
%     \includegraphics[scale=0.5]{figures/num_analysis/ellipse/numerical_analysis_param=theta.pdf}
%     \caption{Caption here}
%     \label{fig:heatmap_training_params_ellipse}
% \end{figure*}

% describe and quantify heatmap plots with 2D function for ellipse size/angle, SEE brightness opening angle

%% Broadly, all parameters show relatively smooth and simple behavior modification with spin and inclination.

The ellipse semimajor and semiminor axes as well as the orientation show substantially less variation across the Kerr parameter space. In contrast to the SEE parameters, the size parameters of the ellipse fit vary $\lesssim6\%$ as spin and inclination are modified. In general, the lower variation makes spin predictions based solely on ellipse fits significantly less discriminatory than predictions based on the SEE model, or a similar model which incorporates brightness distribution.

\begin{figure*}[]
    \centering
    \includegraphics[scale=0.55]{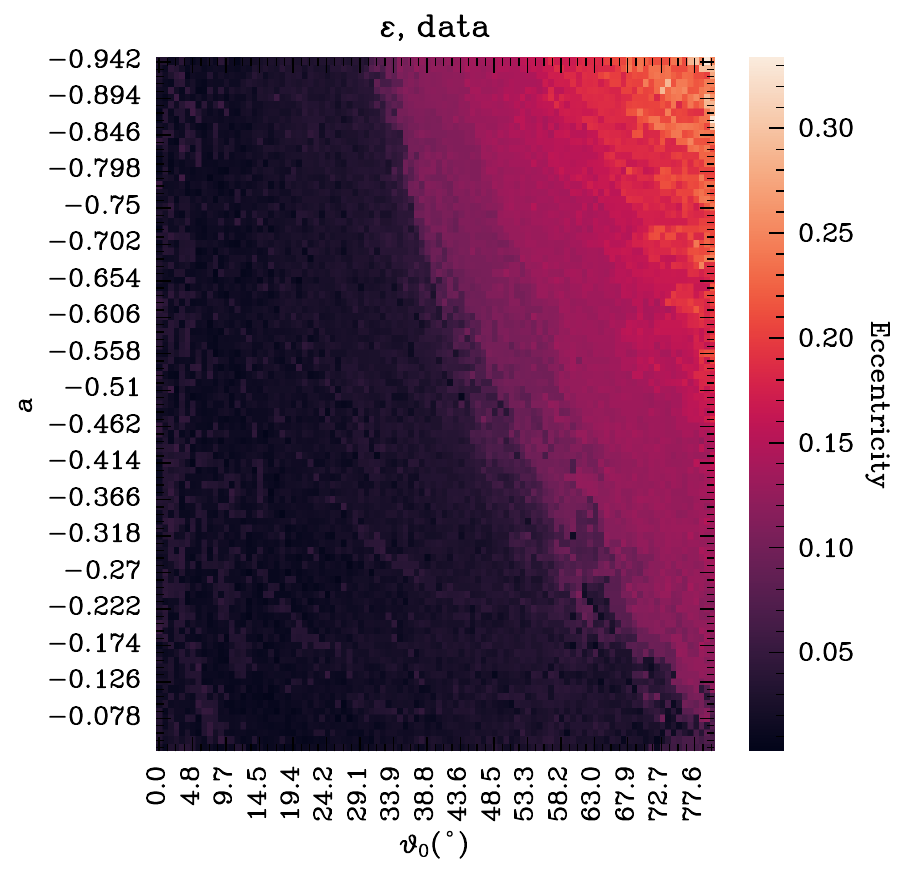}
    \includegraphics[scale=0.55]{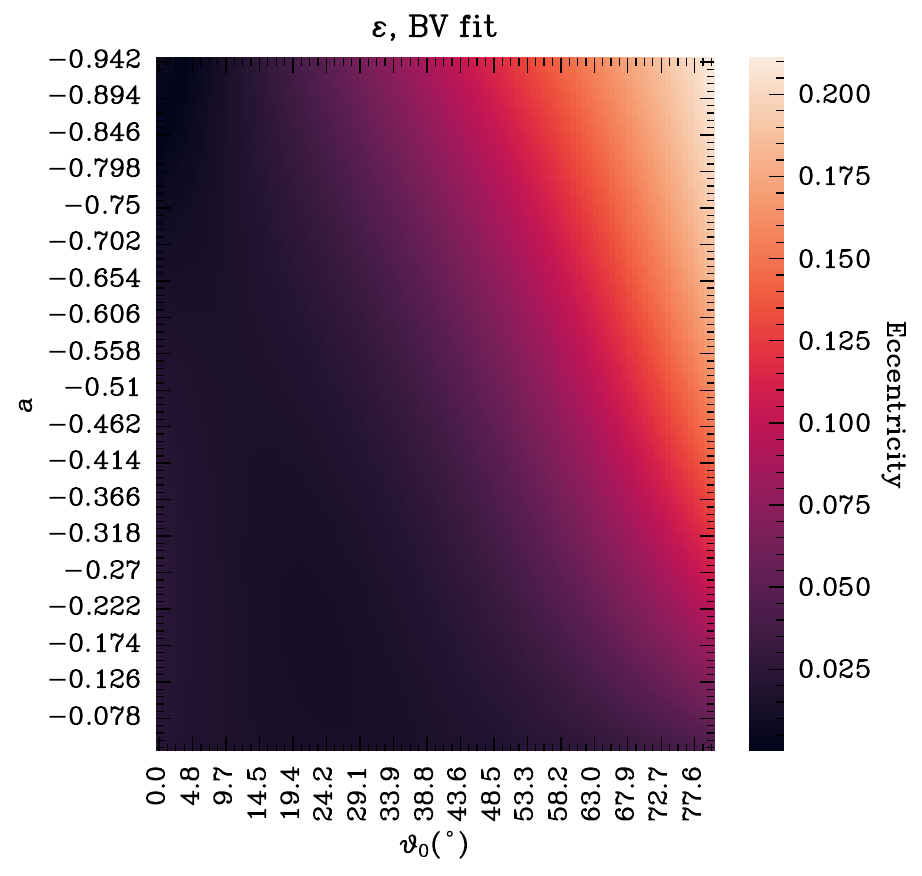}
    \caption{Example visualization of the fits to the SEE and ellipse parameters using a third-order bivariate polynomial (BV). \textit{(left)} The best-fit values of $r_b$ for every combination of spin and inclination in the dataset. \textit{(right)} A description of the $r_b$ parameter using a third-order BV fit. The fit has an average residual of $\lesssim2\%$, indicating a good fit.}
    \label{fig:bestfit_rb}
\end{figure*}

%% We present approximate formulae for the variation of the geometric measurables with spin and inclination. 
The variation for all SEE parameters is more easily characterized by considering only positive or negative spin. To this end, we present approximate formulae for the variation of the geometric measurables with spin and inclination. We fit a third-order bivariate polynomial (BV) $\mathcal{Q}$ in $a$ and $\theta_0$ to an arbitrary quantity $Q(a,\theta_0)$ using the following expression:
\begin{equation}
    \mathcal{Q} = \sum_{i,j=0}^3 A_{i,j} a^i \theta_0^j,
    \label{eq:quantity}
\end{equation}
where $\{A_{i,j}\}$ are the fit coefficients. These fits are performed separately for positive and negative spins. We perform a least-squares minimization over the heatmap distributions in \autoref{fig:heatmap_training_params} using this function and use the best-fit bivariate polynomial to approximate the behavior of the $n=1$ photon ring morphology. An example visualization of the fit to the $r_b$ parameter of the SEE model is shown in \autoref{fig:bestfit_rb}. We present BV parameters for all SEE model parameters in  \autoref{tab:bv_params_negative} and \autoref{tab:bv_params_positive}. On average, we find an average residual of $\lesssim2\%$ across all spins and inclinations, which indicates excellent fits for all parameters. 

\begin{table*}
    \begin{centering}
    \begin{tabular}{cccccccccc}
    % \bottomrule
    $a < 0$  & & & & & & & & \\
    \bottomrule
    \textbf{Parameter} & $A_{00}$ & $A_{01}$ & $A_{02}$ & $A_{10}$ & $A_{11}$ & $A_{12}$ & $A_{20}$ & $A_{21}$ & $A_{22}$ \\ \midrule
    % $r_b$ &  $5.800$  &  $-0.013$  &  $2.78\times10^{-4}$  &  $2.450$  &  $-0.309$  &  $0.005$  &  $2.950$  &  $-0.351$  &  $0.005$ \\ \cdashline{2-10}
    % $r_a$ &  $6.000$  &  $-0.019$  &  $2.52\times10^{-4}$  &  $0.369$  &  $0.002$  &  $-8.17\times10^{-4}$  &  $-0.921$  &  $0.122$  &  $-0.003$ \\ \cdashline{2-10}
    $\varepsilon$ &  $0.206$  &  $-0.008$  &  $  1.0\times10^{-4}$  &  $-0.002$  &  $0.023$  &  $ -7.6\times10^{-4}$  &  $-0.412$  &  $0.065$  &  $-0.001$ \\ \cdashline{1-10}
    $|\vec{C}_{\mathrm{m}} - \vec{C}_{\mathrm{S}}|^x/r_a$ & $-0.062$  &  $0.006$  &  $ -9.3\times10^{-5}$  &  $-0.338$  &  $0.031$  &  $ -3.5\times10^{-4}$  &  $-0.248$  &  $0.019$  &  $-1.8\times10^{-4}$ \\ \cdashline{1-10}
    $|\vec{C}_{\mathrm{m}} - \vec{C}_{\mathrm{S}}|^y/r_a$ &  $-0.008$  &  $  5.5\times10^{-4}$  &  $ -2.0\times10^{-5}$  &  $-0.081$  &  $0.010$  &  $ -1.5\times10^{-4}$  &  $-0.080$  &  $0.010$  &  $-1.2\times10^{-4}$ \\ \cdashline{1-10}
    $\varphi_{\mathrm{S}}-\arg(\vec{C}_{\mathrm{m}} - \vec{C}_{\mathrm{S}})$ &  $0.796$  &  $-3.400$  &  $0.036$  &  $150.000$  &  $-8.830$  &  $0.123$  &  $133.000$  &  $-8.790$  &  $0.123$ \\ \cdashline{2-10}
    \toprule
    \end{tabular}
    \caption{BV fit values for the $A_{ij}$ matrix components for negative spin ($a < 0$), computed for each SEE model parameter, as constructed in \autoref{sec:ellipse}.}
    \label{tab:bv_params_negative}
    \end{centering}
\end{table*}

\begin{table*}
    \begin{centering}
    \begin{tabular}{cccccccccc}
    % \bottomrule
    $a > 0$  & & & & & & & \\
    \bottomrule
    \textbf{Parameter} & $A_{00}$ & $A_{01}$ & $A_{02}$ & $A_{10}$ & $A_{11}$ & $A_{12}$ & $A_{20}$ & $A_{21}$ & $A_{22}$ \\ \midrule
    % $r_b$ &  $5.800$  &  $-0.013$  &  $2.78\times10^{-4}$  &  $2.450$  &  $-0.309$  &  $0.005$  &  $2.950$  &  $-0.351$  &  $0.005$ \\ \cdashline{2-10}
    % $r_a$ &  $6.000$  &  $-0.019$  &  $2.52\times10^{-4}$  &  $0.369$  &  $0.002$  &  $-8.17\times10^{-4}$  &  $-0.921$  &  $0.122$  &  $-0.003$ \\ \cdashline{2-10}
    $\varepsilon$ &  $0.011$  &  $2.5\times10^{-4}$  &  $-8.1\times10^{-7}$  &  $-0.002$  &  $2.8\times10^{-4}$  &  $1.0\times10^{-6}$  &  $0.006$  &  $5.4\times10^{-4}$  &  $-7.1\times10^{-6}$ \\ \cdashline{1-10}
    shfitx &  $8.0\times10^{-4}$  &  $-0.003$  &  $3.2\times10^{-5}$  &  $0.004$  &  $0.002$  &  $-4.3\times10^{-5}$  &  $-0.019$  &  $0.004$  &  $-1.6\times10^{-5}$ \\ \cdashline{1-10}
    shifty &  $0.014$  &  $8.0\times10^{-4}$  &  $-9.2\times10^{-7}$  &  $-0.055$  &  $0.007$  &  $-9.3\times10^{-5}$  &  $0.059$  &  $-0.008$  &  $1.1\times10^{-4}$ \\ \cdashline{1-10}
    alpha &  $-23.700$  &  $-3.610$  &  $0.044$  &  $24.100$  &  $-6.140$  &  $0.097$  &  $-10.200$  &  $3.660$  &  $-0.061$ \\ \cdashline{1-10}
    \toprule
    \end{tabular}
    \caption{BV fit values for the $A_{ij}$ matrix components for positive spin ($a > 0$), computed for each SEE model parameter, as constructed in \autoref{sec:ellipse}.}
    \label{tab:bv_params_positive}
    \end{centering}
\end{table*}

% \clearpage
\section{Machine learning application}
\label{sec:fitting}

% \subsection{Choice of machine-learning algorithm}

% summarize neural networks
Recent advancements in machine learning have revolutionized various scientific domains, including astrophysics, by enabling the development of sophisticated data analysis techniques. Among these, gradient boosting and convolutional neural networks (CNNs) have emerged as particularly impactful. Gradient boosting \cite{GradientBoosting,XGBoost} has proven effective for structured data analysis, offering robust predictive performance through the ensemble of weak learners. Concurrently, CNNs \citep{CNN1,CNN2} have significantly advanced computer vision tasks, providing powerful tools for image recognition and classification. Here, we apply these technques to the study of the $n=1$ photon ring. In \autoref{sec:deep_horizon}, we implement a convolutional neural network following the Bayesian neural network in \cite{Jordy} to estimate spin $a$ directly from images of the $n=1$ photon ring. In addition, we attempt to approximate the anticipated performance of such a neural network by training a gradient boosting regression model on the geometric measurables presented in \autoref{sec:ellipse}.

\subsection{Setup and training}
\label{sec:gbrt}

We implement a gradient boosting regression tree (GBRT) in \texttt{scikit-learn} following the notation of \cite{scikit-learn} and the references therein in order to estimate spin from the geometric measurables presented in \autoref{sec:ellipse}. We briefly summarize the techniques used here. Gradient boosting is a machine learning technique that incrementally builds an ensemble of weak predictive models, typically decision trees, to optimize performance by minimizing a loss function through iterative improvements. A GBRT regressor constructs a prediction $\hat{y}_i$ for a given input $x_i$ by adding the predictions of a series of $M$ weak ``learners'' $h_m{x_i}$; i.e., 
\begin{equation}
    \hat{y}_i = \sum_{m=1}^M h_m(x_i).
\end{equation}
Each estimator $h_m$ is a decision tree regressor of fixed size, constructed to minimize the losses of the summed previous $m-1$ estimators. The initial estimator $h_0$ is arbitrary and, for a least-squares loss function, is typically chosen as the mean of the response variables. The parameters of the trees are fit via a functional space gradient descent.

We implement the \texttt{GradientBoostingRegressor} in \texttt{scikit-learn} initialized with $M=125$ and a least-squares loss function. The value of $M$ was chosen based on performance on the validation set. As $M$ increases, the performance of the network asymptotically approaches some maximum accuracy, and can even decrease as overfitting effects increase. We choose the number of learners $M$ such that the accuracy achieves $\approx99\%$ of the maximum, to maximize performance while avoiding overfitting. We train on the input quantities $(r_a, r_b, |\vec{C}_{\mathrm{m}} - \vec{C}_{\mathrm{S}}|, \varphi_{\mathrm{S}}-\arg(\vec{C}_{\mathrm{m}} - \vec{C}_{\mathrm{S}}))$ described in \autoref{sec:ellipse} and response variable spin ($a$). We train on 90\% of the full ($10^6$ images) dataset, separating 10\% ($10^5$ images) of the dataset as a universal validation dataset (which the network never sees) to prevent overfitting. For testing purposes, we also attempted training with 95\% and 80\% of the full dataset (with 5\% and 20\% of the dataset for validation, respectively) and did not notice a significant variation ($\lesssim 1\%$) in the performance of the algorithm. The raw prediction of the algorithm $\hat{y}_i$ is treated as the spin estimate with no modifications.

We next estimate an uncertainty on our measurement. Broadly, there are two types of estimate uncertainties: \textit{epistemic} uncertainties, capturing the insuffiency of the model; and \textit{aleatoric} uncertainties, which capture the intrinsic uncertainty present in the data. We discuss both below. 

To estimate an epistemic uncertainty, we implement cross-validation similar to the stochastic gradient boosting technique \citep{StocGradBoost}.
In addition to the main prediction algorithm, we perform 10 leave-one-out bootstrapping steps, which we refer to as ``bootstrap samples''. Each training set of these samples is randomly selected to be 40\% of the whole dataset, with the remaining 10\% used as the validation set. After an estimate of spin is made by using the whole dataset for a prediction, the samples are similarly queried to produce a range of predictions $\{\hat{y}_k\}_{k=0}^{10}$ based on their limited sample of the dataset. The standard deviation of this range $\sigma_{\hat{y},k}$ is used as an estimate of the epistemic uncertainty. The purpose of this uncertainty is to estimate how sensitive our prediction is to the particular construction of the dataset and choice of input images. 

To estimate an aleatoric uncertainty, we consider the largest source of uncertainty in an estimate of spin: namely, the degeneracy of the photon ring shape as the inclination varies. At low inclination, the geometric shape of the photon ring is virtually invariant to spin. The degeneracies are visible in \autoref{fig:heatmap_training_params}, where most parameters see little-to-no variation at inclinations $\theta_0\lesssim40^\circ$. To quantify the uncertainty due to this degeneracy, for each spin $a$ we consider a range of inclinations $[\theta_0, \theta_0+\delta\theta_0]$. Within this range, we adopt the standard deviation $\sigma_{\theta_0}$ of the estimates of spin for each $(a,\theta_0)$ combination as an estimate of the aleatoric uncertainty resulting from the inclination degeneracy. We choose $\delta\theta_0=19.75^\circ$, corresponding to 25\% of the inclination range we generated for the dataset. Each bin of inclination would therefore contain $\approx2.5\times10^3$ spin estimates at that inclination range, making it an effective approximation of the estimate variance due to the inclination degeneracy. 

We add together our aleatoric and epistemic uncertainties together in quadrature to construct a global uncertainty estimate. In general, we found that $\sigma_{\hat{y},k} \ll \sigma_{\theta_0}$, making the inclination degeneracy the dominant source of uncertainty in our estimates.

\subsection{Overall performance of the GBRT}
% FIGURES: (1) histogram of overall performance, (2) model of performance with spin and inclination

\begin{figure*}
    \centering
    \hspace{0.25cm}
    \includegraphics[scale=0.4]{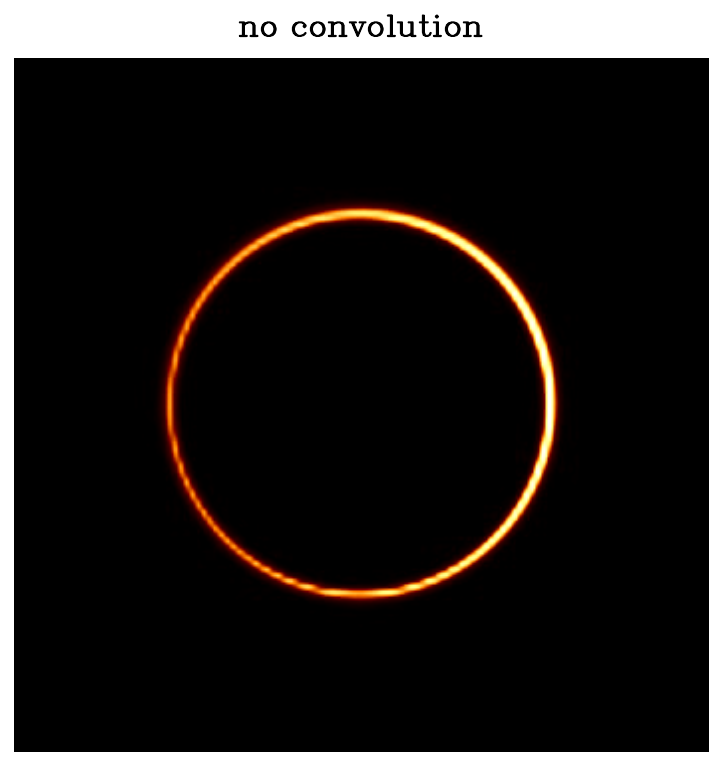}
    \hspace{0.25cm}
    \includegraphics[scale=0.4]{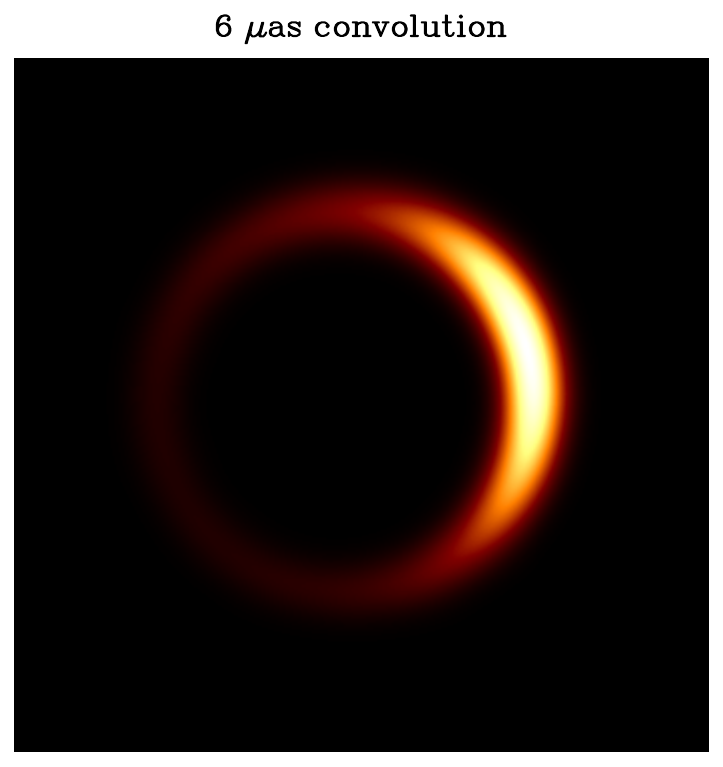}
    \hspace{0.25cm}
    \includegraphics[scale=0.4]{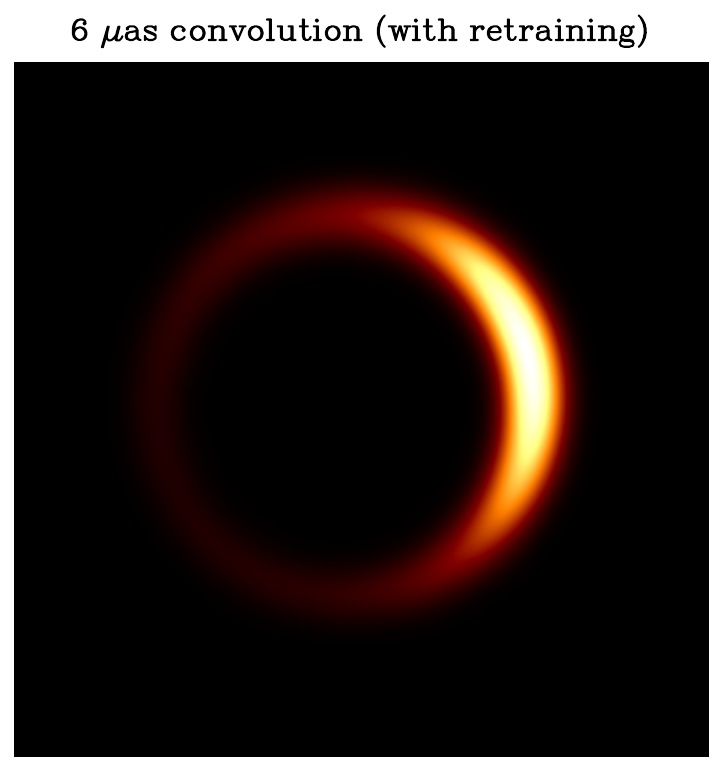}
    \includegraphics[scale=0.4]{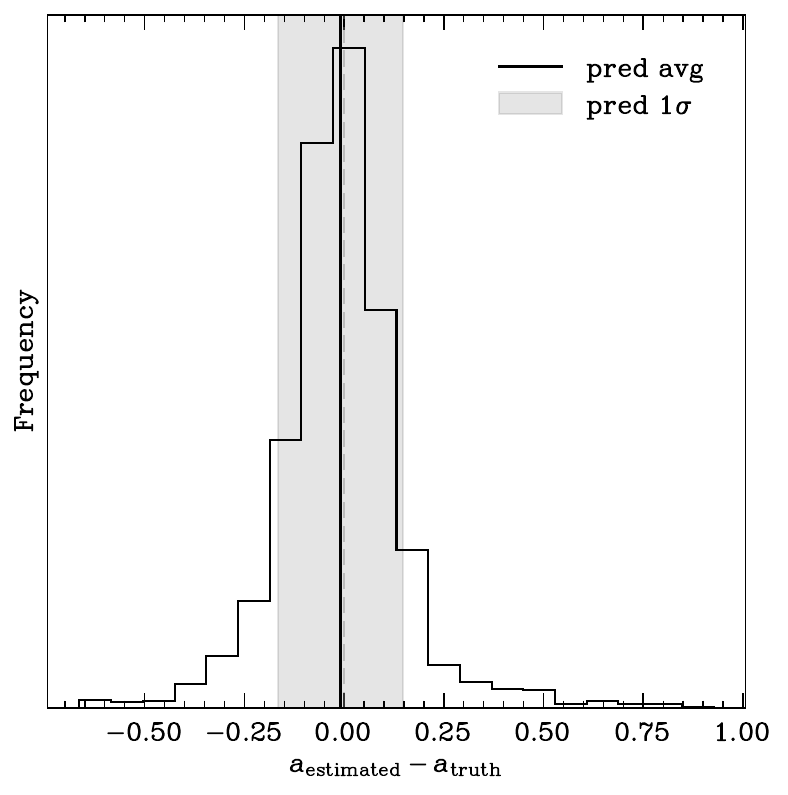}
    \includegraphics[scale=0.4]{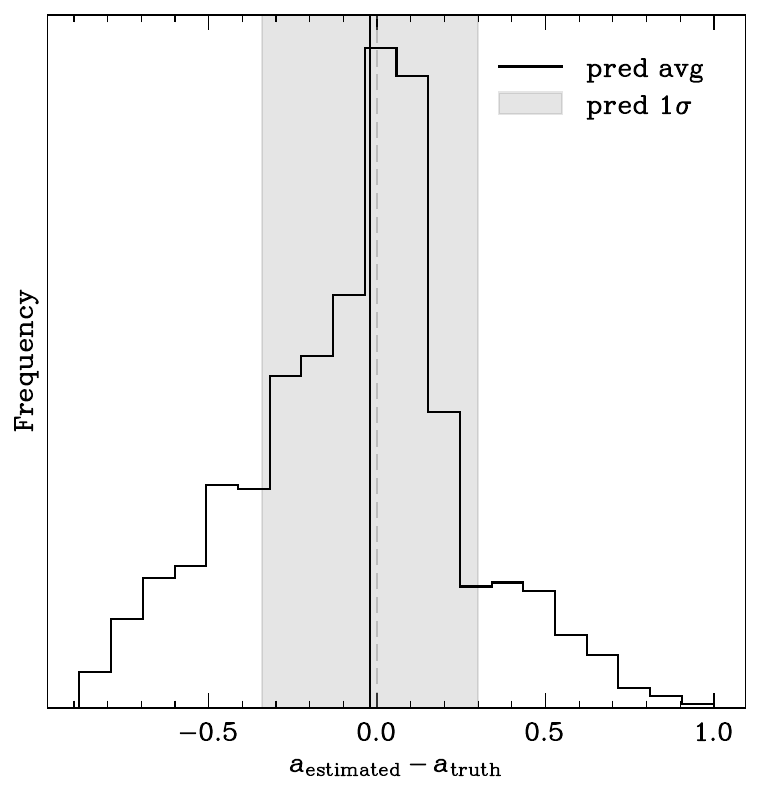}
    \includegraphics[scale=0.4]{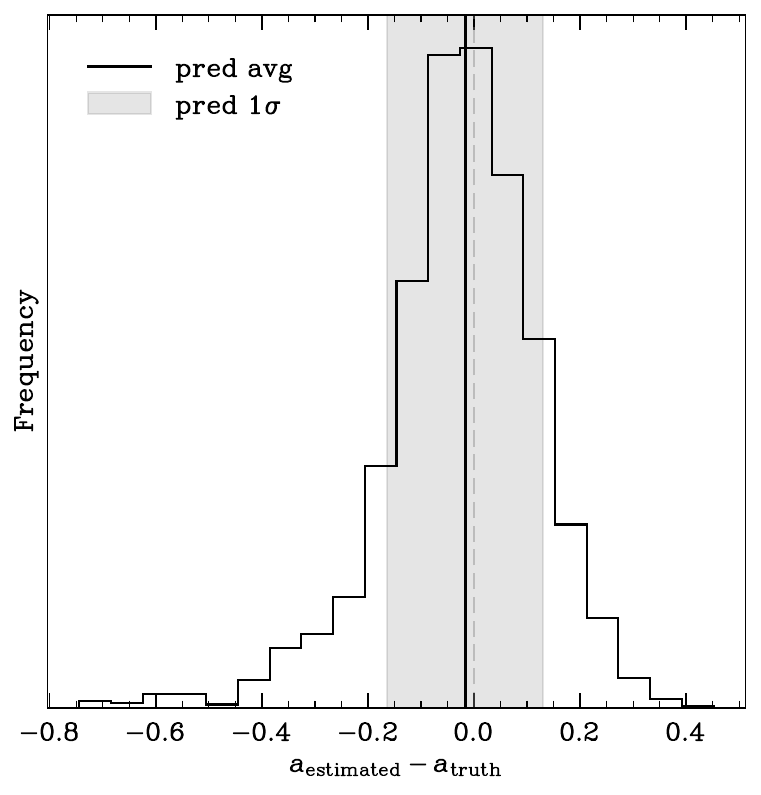}
    \\
    % \hspace{10.5cm}\includegraphics[scale=0.4]{figures/overall_performance/spin_deviation_global_BLUR20_TRAINED.pdf}
    \caption{Performance at varying convolution for the GBRT described in \autoref{sec:gbrt}. \textit{(top)} Representative appearances of the $n=1$ photon ring at varying convolutions, ranging from no convolution (left) to the anticipated BHEX beam at 345 GHz (center and right). \textit{(bottom)} Performance of the GBRT with the convolution level shown in the top row. With no convolution, or convolution with retraining, recovery is highly accurate. As the size of the convolving beam is increased, the performance of the network without retraining predictably deteriorates.}
    \label{fig:overall_residuals_spin}
\end{figure*}

We consider the overall performance of the GBRT described in \autoref{sec:gbrt} with and without convolution, and with retraining. The maximum nominal BHEX beam is expected to be $\approx9$ \muas, corresponding to operations at 230 GHz, and resulting in an imaging resolution of $\approx4.5$ \muas. At 345 GHz, the BHEX beam is expected to be $\approx6$ \muas, corresponding to an imaging resolution of $\approx3$ \muas. Thus, to conservatively simulate BHEX imaging conditions, we test the performance of the GBRT under a convolution of $6$ \muas, with and without retraining. The distribution of residuals between the predicted and estimated spins is shown in \autoref{fig:overall_residuals_spin}. We find an average residual of $\approx0$ ($r^2 \approx0.95$) with a spread of $\lesssim\pm0.2$ on the unconvolved dataset. This corresponds to $\lesssim20\%$ of the full range of spins and is convincing evidence that, in the absence of noise and other artifacts, the photon ring geometry and brightness distribution is sufficient to distinguish between different spins. The performance is significantly worsened after convolution with a 6$\mu$as beam. At this convolution, the average residual is still $\approx0$; however, the spread of deviations is $2\text{-}3$x larger than the spread of deviations with no convolution. Additionally, there is a skew towards negative values of $a_{\mathrm{estimated}}-a_{\mathrm{truth}}$, indicating that the effect of convolution on our pipeline is an underestimate of spin for a significant portion of the dataset. 

\subsection{Intra-inclination performance of the GBRT}

\begin{figure*}
    \centering
    \includegraphics[scale=0.412]{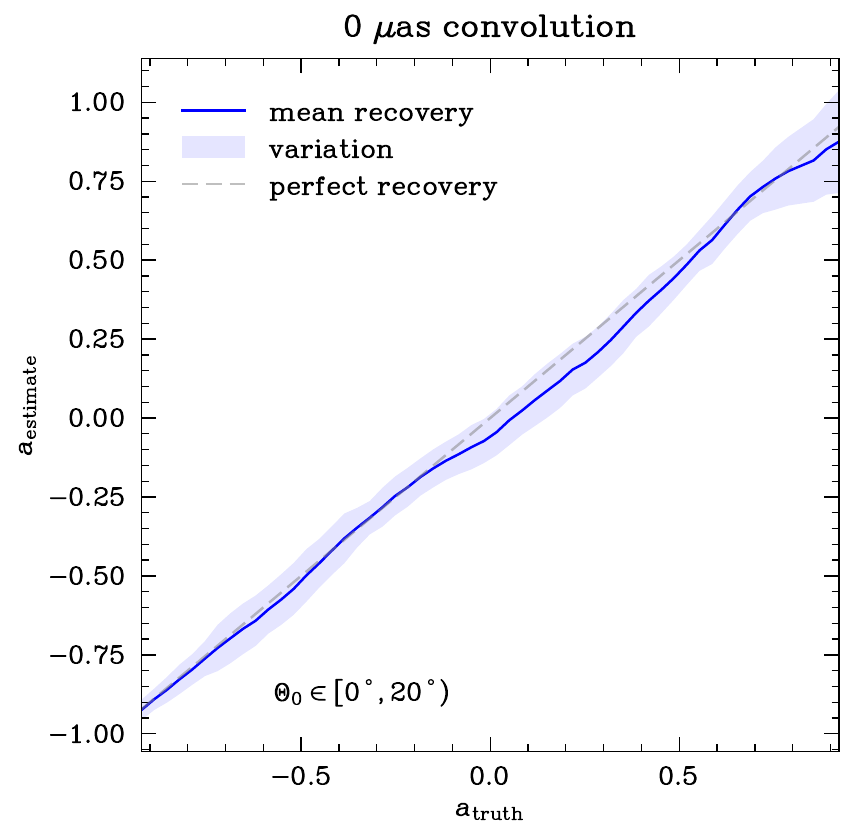}
    \includegraphics[scale=0.412]{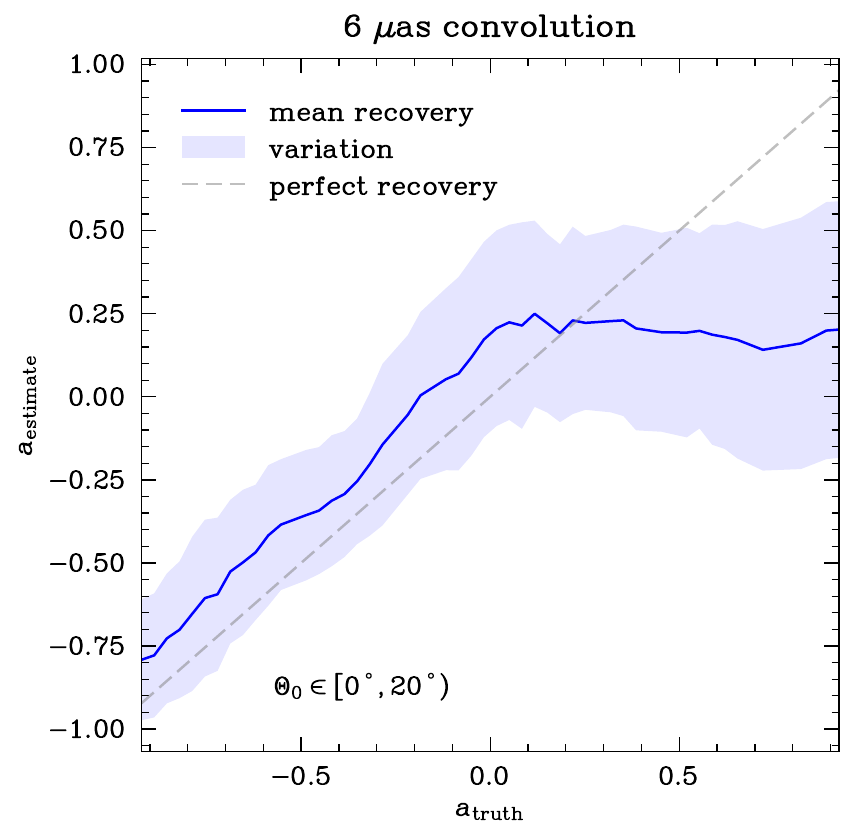}
    \includegraphics[scale=0.412]{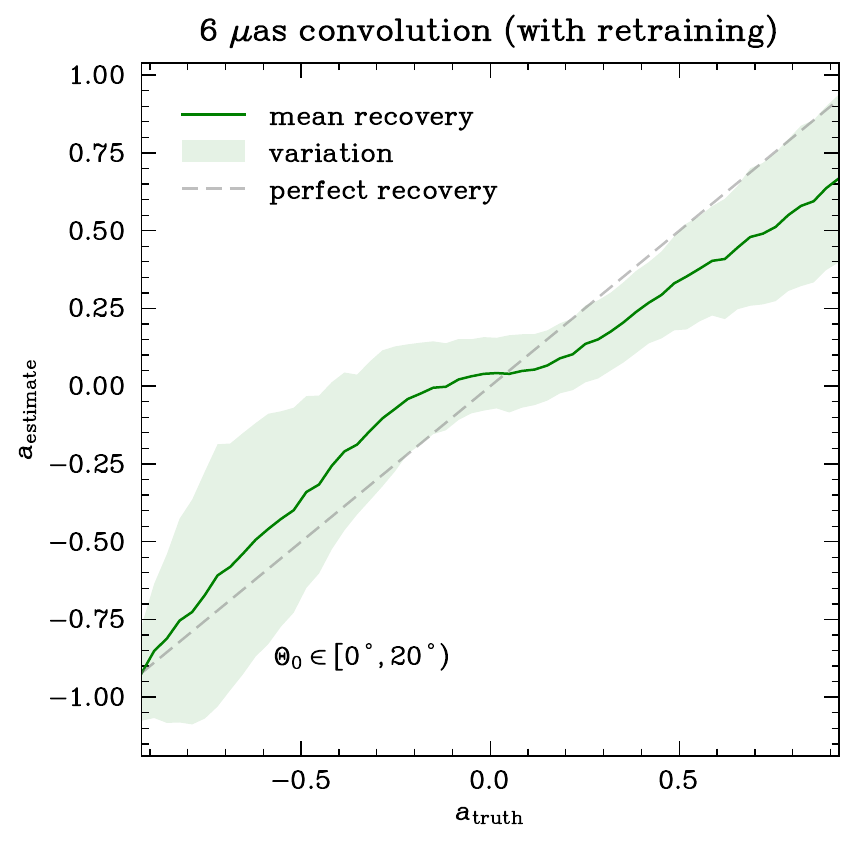}

    \includegraphics[scale=0.412]{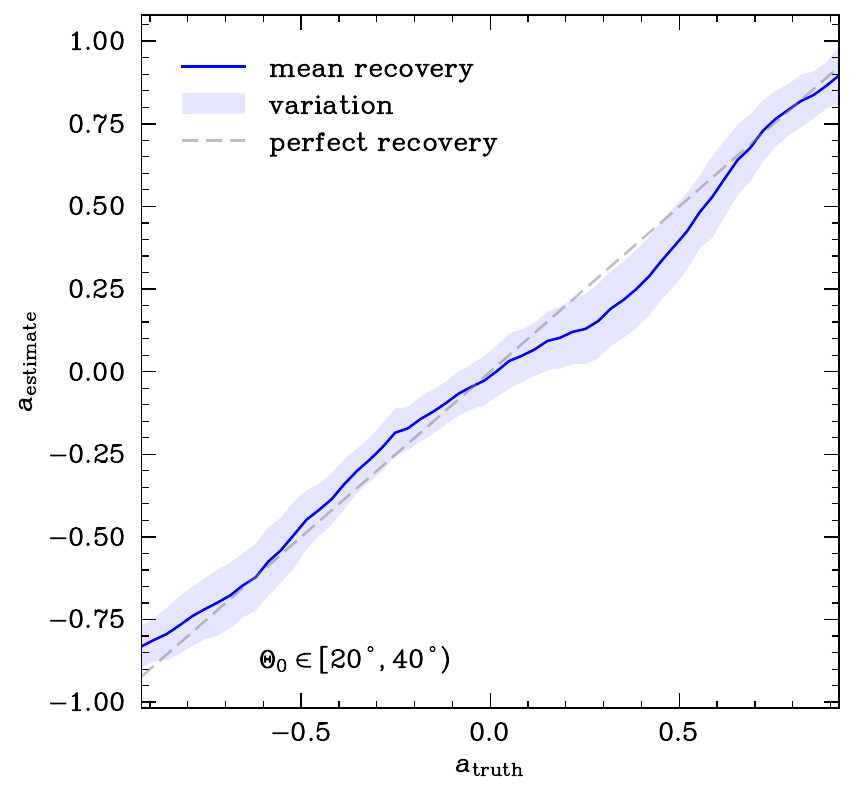}
    \includegraphics[scale=0.412]{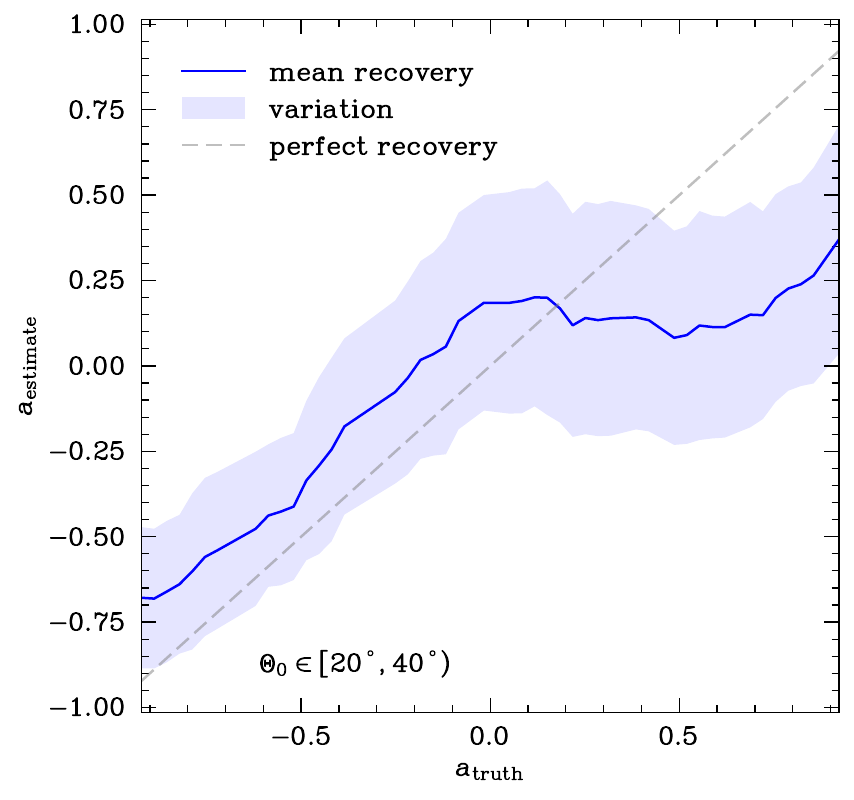}
    \includegraphics[scale=0.412]{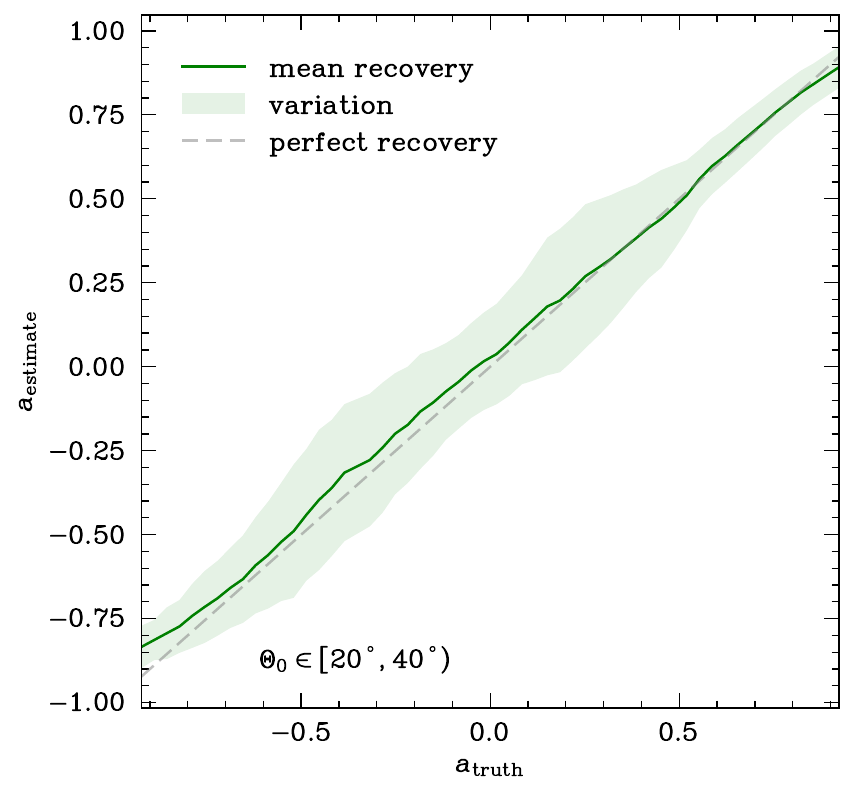}

    \includegraphics[scale=0.412]{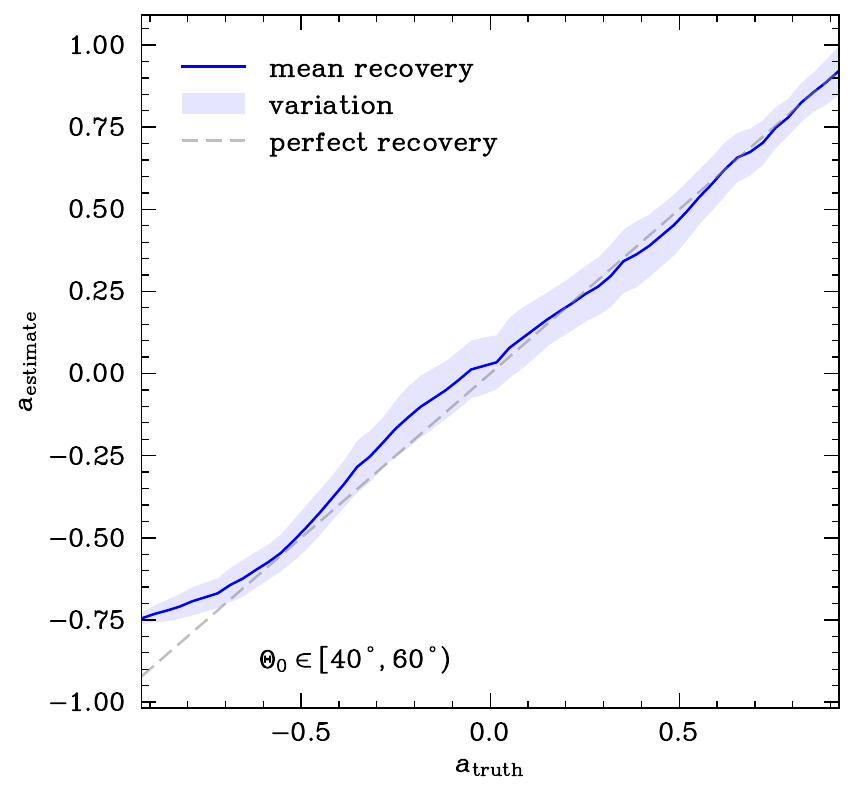}
    \includegraphics[scale=0.412]{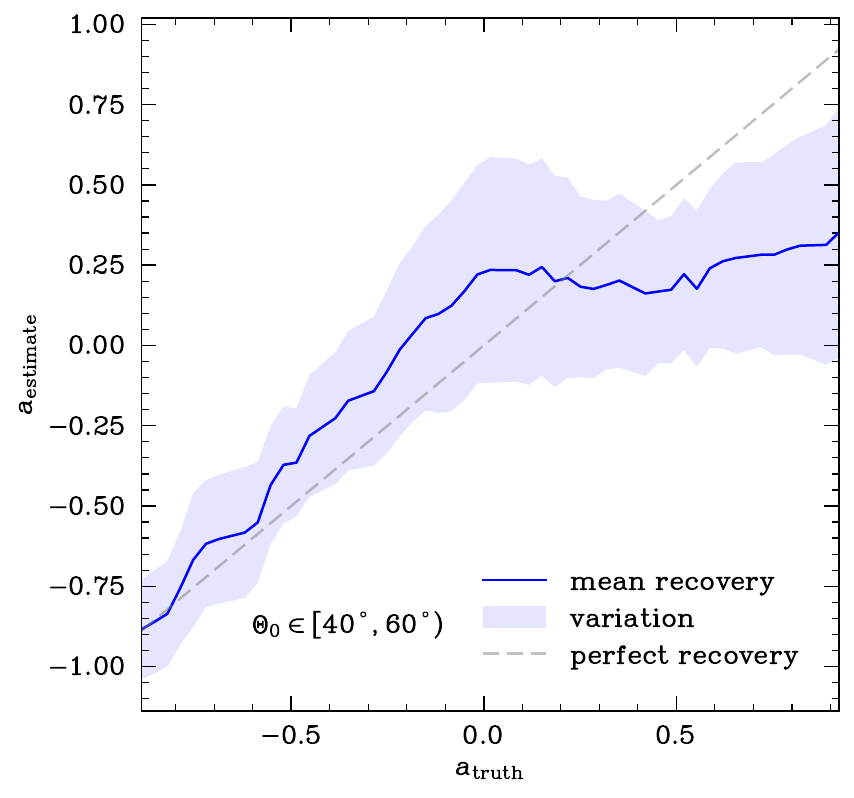}
    \includegraphics[scale=0.412]{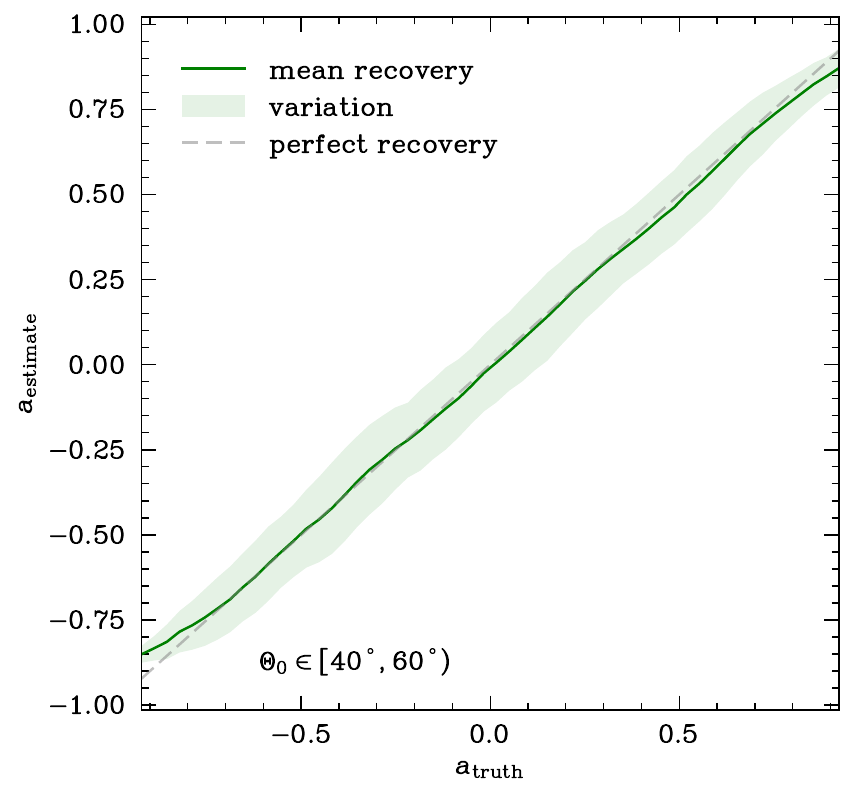}

    \includegraphics[scale=0.412]{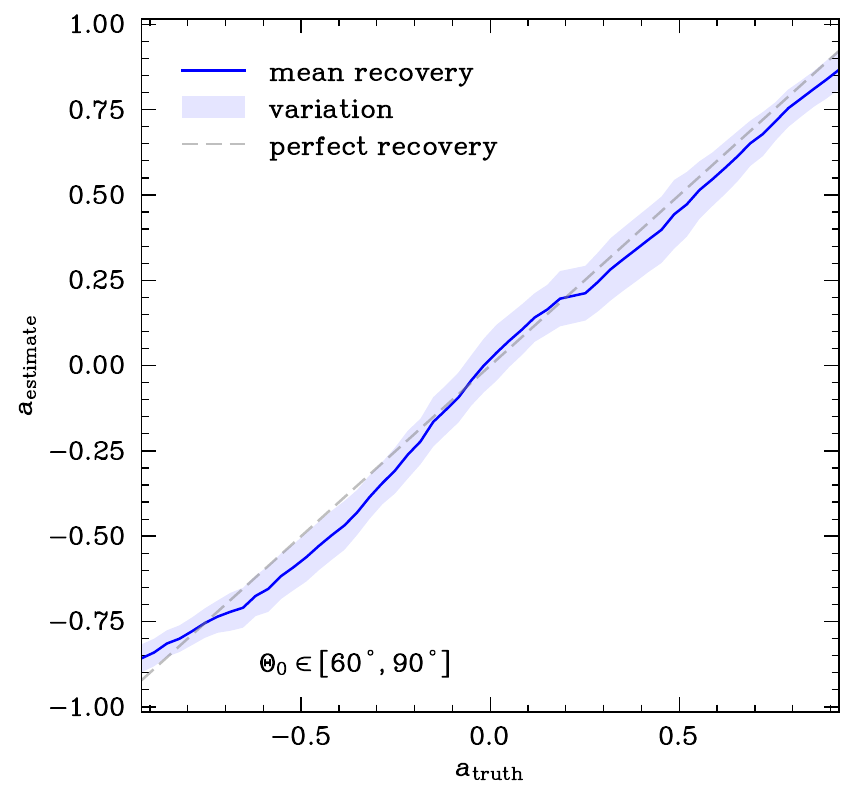}
    \includegraphics[scale=0.412]{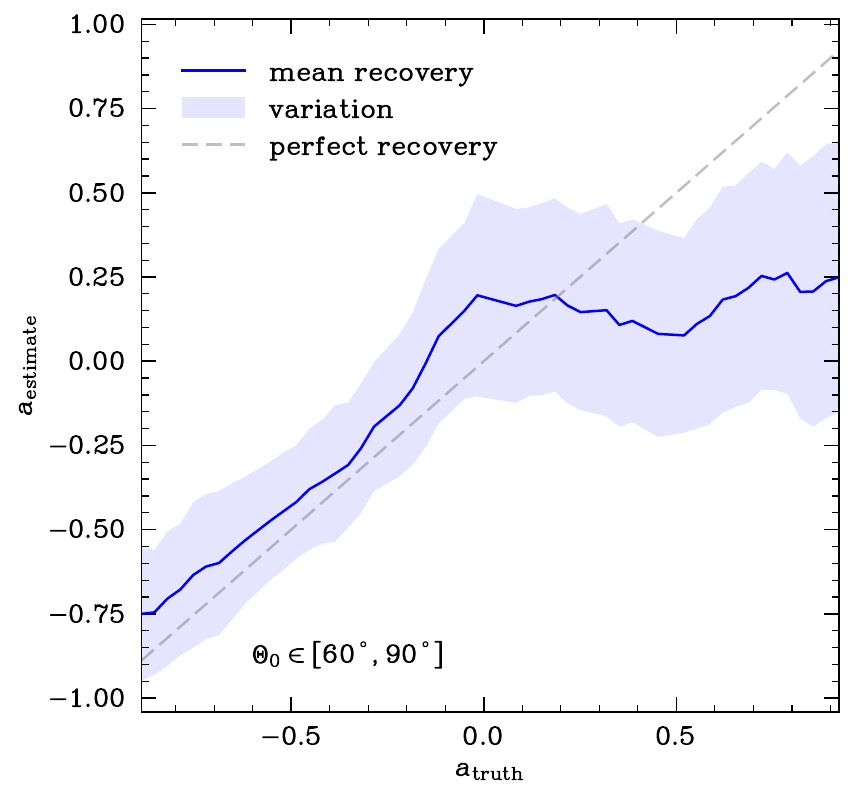}
    \includegraphics[scale=0.412]{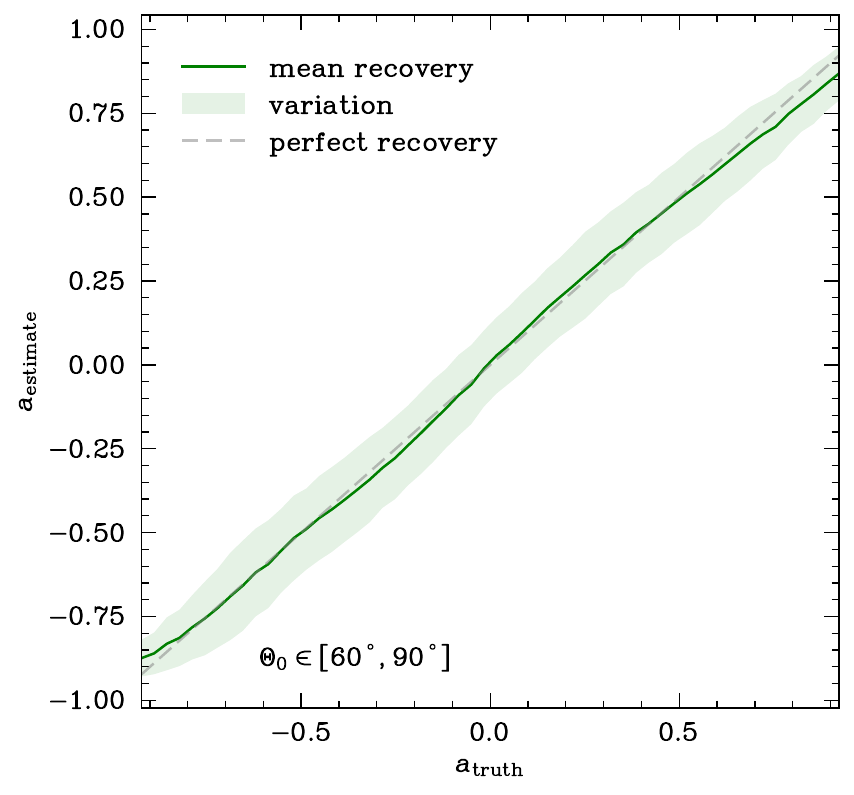}

    \caption{Intra-inclination performance of the GBRT on four inclination ranges: face-on ($0^\circ$ to $20^\circ$), low ($20^\circ$ to $40^\circ$), high ($40^\circ$ to $60^\circ$), and edge-on ($60^\circ$ to $90^\circ$). The left column (blue) corresponds to the GBRT trained on non-convolved data operating on non-convolved data; the right column (green) is the equivalent analysis for the GBRT operating on data convolved with the $6$ \muas beam. The center column corresponds to the GBRT trained on non-convolved data operating on data convolved with a 6 $\mu$as BHEX beam. The shaded region in all plots corresponds to the aleatoric uncertainty $\sigma_{\theta_0}$ computed in \autoref{sec:gbrt}. }
    \label{fig:inc_performance}
\end{figure*}

We now consider the performance of the GBRT at different inclinations. Due to the inclination degeneracy, we expect variable performance of the neural network as inclination is varied. In \autoref{fig:inc_performance}, we visualize the performance of the GBRT on four inclination ranges: face-on ($0^\circ$ to $20^\circ$), low ($20^\circ$ to $40^\circ$), high ($40^\circ$ to $60^\circ$), and edge-on ($60^\circ$ to $90^\circ$). We perform this visualization for the GBRT trained on non-convolved data operating on non-convolved data, as well as the equivalent analysis for the GBRT operating on data convolved with a $6$ \muas beam. Unexpectedly, we find that GBRT performance at low and face-on inclinations in the non-convolved case is similar to the performance at high and edge-on inclinations. Recovery at all inclinations is consistent (within the aleatoric uncertainty) with perfect recovery, although there is a systematic under-estimation of the spin magnitude of $\lesssim5\%$ (of the full spin range) at spins with $a < -0.8$ for $\theta_0\in[40^\circ, 60^\circ)$. The other inclination ranges show comparable performance. The excellent performance even at low spins and inclinations indicates that, in the absence of imaging convolution and artifacts, the $n=1$ photon ring can be a distinguishing proxy for spin. 

The network performance is significantly worsened when realistic BHEX convolution is introduced in images. When images are convolved with a $6$ \muas beam (no retraining), negative spins are recovered accurately (though with a significantly larger spread in predictions). However, positive spins are largely under-estimated, with most estimates being consistent with $a=0$. When the network is retrained on images convolved with the $6$ \muas BHEX beam, the performance is comparable to the $0$ \muas convolution test at inclinations $\theta_0 > 20^\circ$. At inclinations $\theta_0 < 20^\circ$, estimates are still consistent with the true values, but there is a greater tendency to underestimate the spin magnitude. 

The lower performance with convolution indicates that an image feature-extraction approach may be inadequate to perform spin extraction using the $n=1$ photon ring unless estimation tools are trained specifically at the anticipated BHEX resolution. However, the excellent performance with no convolution indicates that the $n=1$ photon ring can be a distinguishing proxy for spin in the absence of convolution or imaging artifacts. Performing spin estimates using Fourier analogs of the geometric measurables may sidestep confounding effects from imaging; we investigate this method of spin inference in \cite{Farah2024ip}.

\subsection{Comparison to Deep Horizon}
\label{sec:deep_horizon}

Convolutional neural networks (CNNs) and the cousin Bayesian neural networks (BNNs) have been used in past analyses to demonstrate recovery of simulation parameters from GRMHD simulations of black hole accretion flows \citep{Jordy}. The primary benefit of a CNN is the ability to train on and input entire images, without needing to perform feature extraction or calculate geometric measurables. This feature of CNNs practically guarantees that they will outperform alternative machine learning algorithms such as GBRT. However, CNNs are also slow to train as a result of running on full, often high-resolution images. By implementing a CNN and comparing the performance to our GBRT, we can assess how effectively our choice of geometric measurables is utilizing the information present in the images.

We reproduce the methodology of \cite{Jordy} with minor modifications. We implement a CNN in \texttt{Tensorflow} version 2.14 \citep{TensorFlow} leveraging the \texttt{keras} \citep{keras} API version 2.14. We choose and implement a 10-layer CNN architecture. Following \cite{Jordy}, we use a rectified linear unit activation function \citep{ReLU} for all layers except the output layer. We use 90\% of the full dataset for training and separate the remaining 10\% for validation. The network is trained on the input data using the legacy implementation of the Adam optimizer \citep{KingmaB14}. The network is initialized with the mean squared error as the primary loss function, with the mean absolute error as an evaluation metric. We use a fixed learning rate of 0.001 for all implementations and runs. 

\begin{figure*}
    \centering
    \includegraphics[scale=0.65]{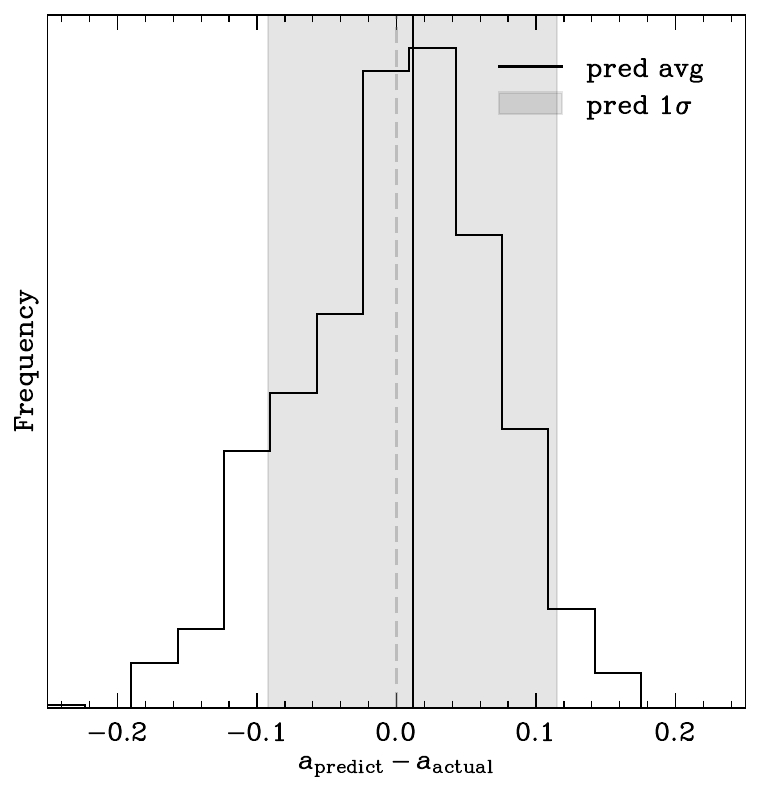}
    \includegraphics[scale=0.65]{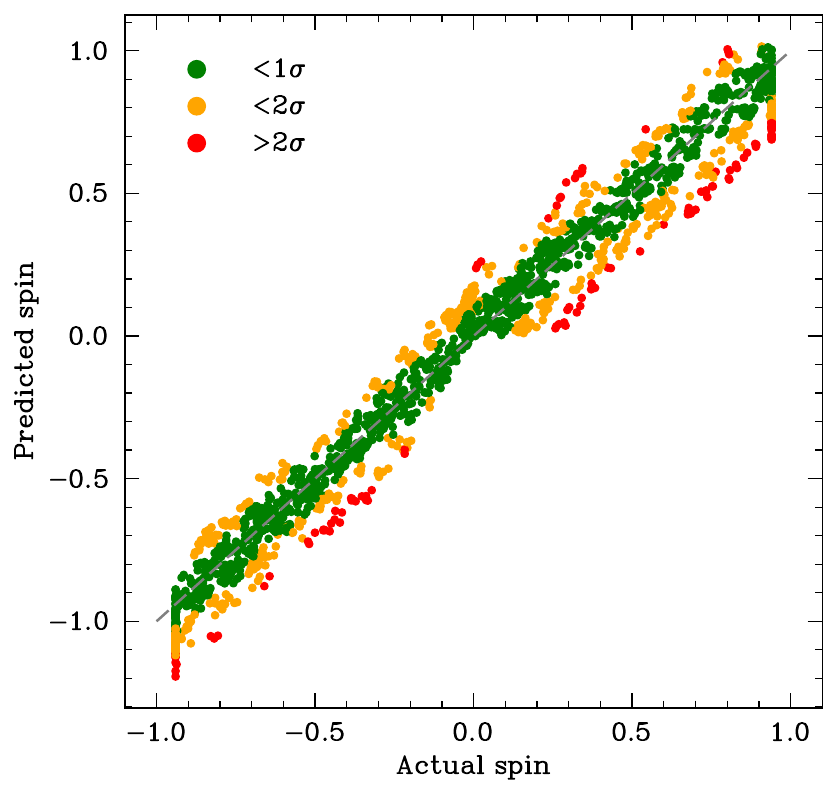}
    \caption{Performance of a CNN trained on non-convolved images. We find recovery performance approximately comparable to the performance of the GBRT. The left histogram is an aggregation of recovery residuals, showing an average residual close to zero with a narrow spread of $\lesssim0.1$. The scatterplot on the right shows sample recoveries at all spins, color-coded by the deviation from the mean recovery at a particular spin.}
    \label{fig:CNN_noconvolved}
\end{figure*}

\begin{figure*}
    \centering
    \includegraphics[scale=0.65]{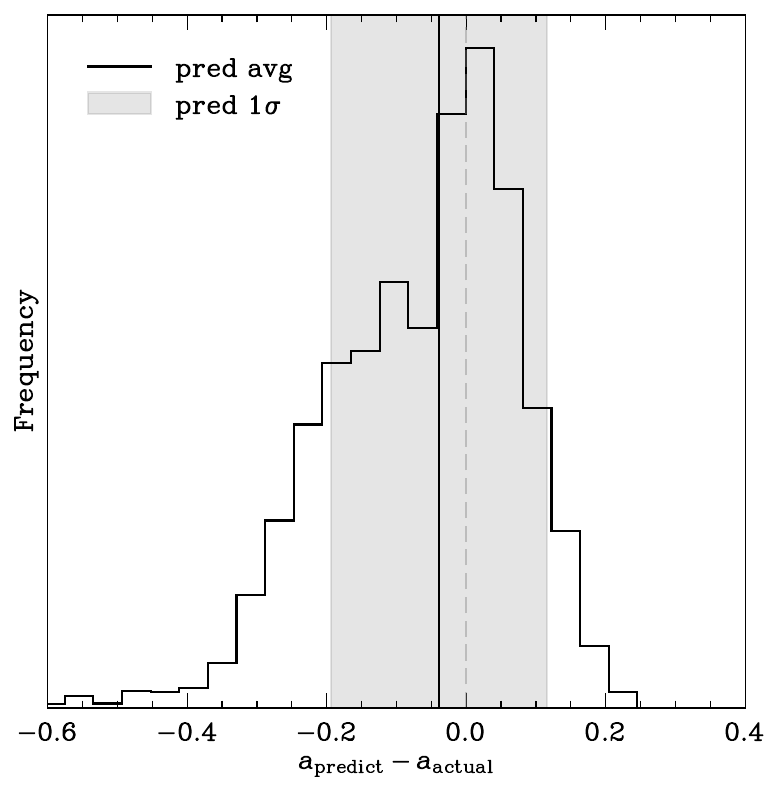}
    \includegraphics[scale=0.65]{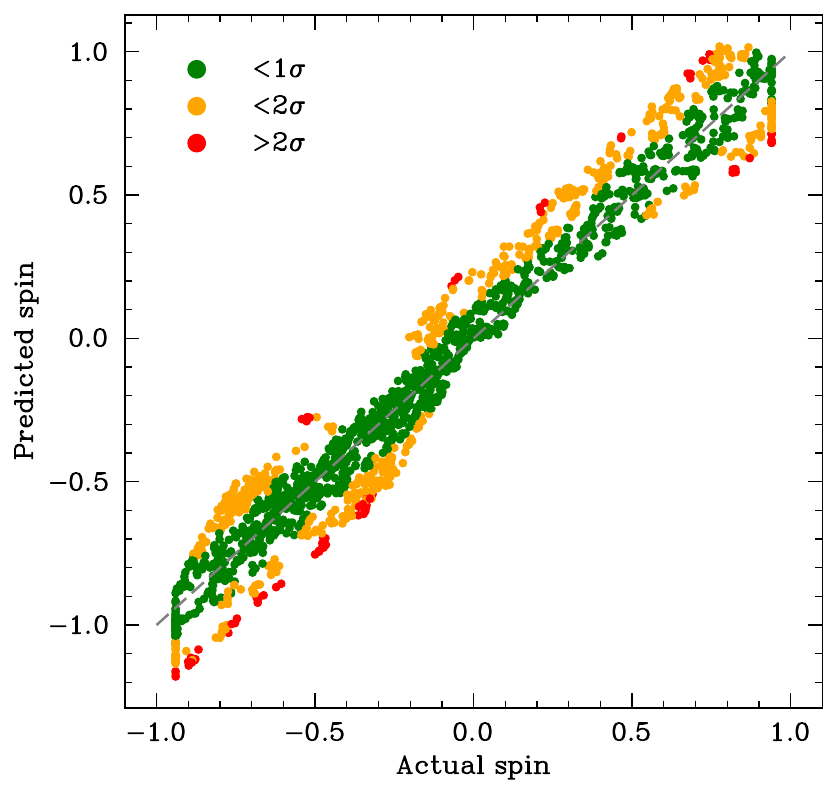}
    \caption{Same as \autoref{fig:CNN_noconvolved}, but for a CNN operating on images convolved with the nominal BHEX beam ($6\mu$as, bottom), without retraining. We find recovery performance approximately comparable to the performance of the GBRT and the non-convolved network. }
    \label{fig:CNN_convolved}
\end{figure*}

We apply the network architecture and training pipeline to both non-convolved images and images convolved with the BHEX beam. The results are shown in \autoref{fig:CNN_noconvolved} and \autoref{fig:CNN_convolved}. The performance on non-convolved images is almost identical to the performance for the corersponding GBRT, featuring a mean residual of $\approx0$ and a $\lesssim20\%$ improvement in the residual spread. We follow \cite{Jordy} and also visualize the performance via a scatter plot of spin recoveries, color-coded by standard deviations from perfect recovery. The network reports consistently accurate recovery of spin at all spin and inclination combinations, with marginally worse performance below $a\lesssim0.2$. The similarity of the performance between the CNN and the GBRT is a strong validation in the choice of geometric measurables being sufficient to capture the salient and spin-distinguishing features of the $n=1$ photon ring. However, the CNN drastically outperforms the GBRT on images convolved with the BHEX beam, indicating the GBRT may not be fully suitable for direct application to BHEX images. 

% \clearpage
\section{Application to GRMHD simulations}
\label{sec:bhex_app}

The analysis previously discussed in this work was based on the semi-analytic simulation engine \texttt{KerrBAM}. Here, we apply the methods developed using \texttt{KerrBAM} to GRMHD simulations of black hole accretion flows. The purpose of this analysis is to provide validation that methods developed using semi-analytic simulation engines can be applied to complex simulations more analogous to a real black hole environment. Additionally, full numerical GRMHD simulations were not included in the training for either the GBRT or CNN networks, making a test on GRMHD simulations a more challenging evaluation of network performance.  

\vspace{1cm}
\subsection{Preparation of synthetic data}
\label{sec:grmhd_synth}

% FIGURES: (1) example simulation and UV coverage and amplitudes vs. baseline length

We analyze the decomposed sub-images from the model set ray-traced in \citet{Palumbo_2022}. These simulations have a mass-to-distance ratio corresponding to the EHT estimate for \virgoa. The fluid models were simulated with {\tt{}iharm3D} \citep{Gammie_HARM_2003, IHARM3d_prather} and ray traced with {\tt{}ipole} \citep{IPOLE_2018}, using the $R_{\rm high}=40$ electron heating parameter described in \citet{Mosci_2016}.  More details about the GRMHD and General Relativstic Ray Tracing (GRRT) pipeline can be found in \citet{Wong_2022}. The resulting images are time-averaged to produce smooth images of typical accretion flow structures, including the photon ring. We generate three GRMHD simulations at $a=0.0, -0.5$ and $-0.94$. The inclinations are generated at the expected inclination angle for \virgoa ($\theta_0\approx17^\circ$).

\subsection{Application to $n=1$ GRMHD simulations}
\label{sec:n1grmhd}
\begin{figure}
    \centering
    \hspace*{-1cm}
    \includegraphics[scale=0.7]{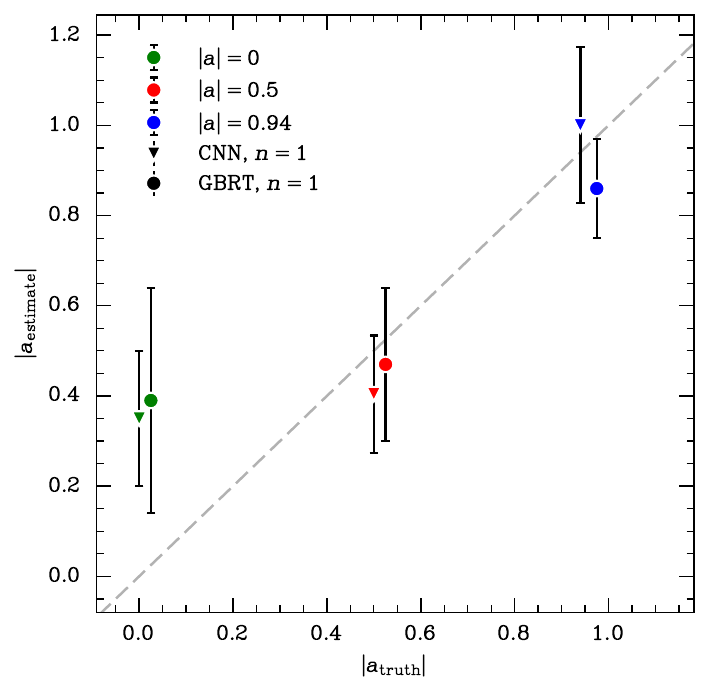}
    \hspace*{-0.3cm}
    \includegraphics[scale=0.6]{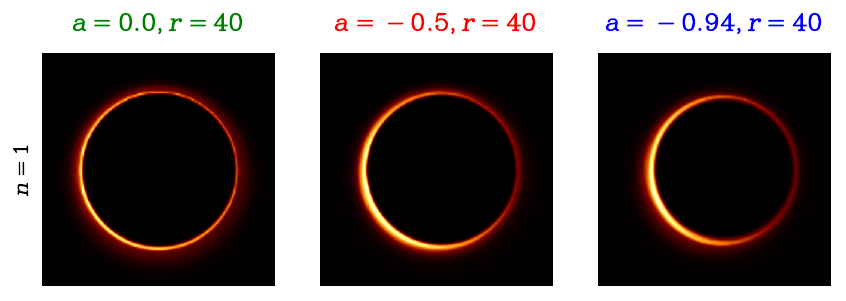}
    \hspace*{-0.3cm}
    \caption{(\textit{top}) Performance of the GRBT (circle points) and CNN (triangle points) on GRMHD images generated following the description in \autoref{sec:grmhd_synth}. The two approaches report very similar performance on GRMHD simulations. Recovery of spin is excellent, despite the low inclination and despite neither using GRMHD images in their training data. Perfect recovery is shown with the gray dashed line. Note: the $|a_\mathrm{truth}|$ values are offset by $\pm0.025$ to improve clarity; the true spin values were either $0.0, -0.5$, or $-0.94$. (\textit{bottom}) The simulated images $n=1$ passed into the GBRT and CNN. }
    \label{fig:grmhd_perf}
\end{figure}

% FIGURES: (1) 
We apply the two algorithms discussed previously (GBRT and CNN) to the GRMHD simulations constructed \autoref{sec:grmhd_synth}. For the GBRT, we measure the geometric observables constructed in \autoref{sec:ellipse} and pass them into the GBRT, which returns an estimate of the spin. For the CNN, we simply resize the image to the 128x128px input size and pass it into the CNN, which returns an estimate of the spin. The results are shown in \autoref{fig:grmhd_perf}. Performance between the two approaches is very similar, despite the difference in method. The similarities in performance on GRMHD are an additional validation that the geometric measurables constructed in \autoref{sec:ellipse} capture most, if not all, of the salient features distinguishing spin. Both methods recovery spin values consistent with the ground truth simulation at $|a|\gtrsim0.5$, but overestimate the low spin simulations in a manner consistent with complications from the spin-inclination degeneracy problem. Performance could be improved by breaking this degeneracy; future methods may incorporate information from e.g., polarization, which is modified by frame-dragging but is also heavily influenced by properties of the plasma flow in the vicinity \citep{Akiyama_2017b,EHT7,Palumbo_2022}; or constraints on the diameter measured in Fourier space, known to vary tightly with spin \citep{Johnson2020,BHEXPRS,Farah2024ip}. 

The accuracy of the spin recovery on both the \texttt{KerrBAM} and GRMHD images is a strong preliminary indication that the $n=1$ photon ring has utility as a proxy to measure spin, with the essential caveat that these tests have been performed on noiseless images without the contribution of the $n=0$ and $n>1$ subring emission. The presence of additional subrings in BHEX images--particularly the $n=0$ direct emission, which is dominated by surrounding astrophysical effects and not the black hole properties--will likely overwhelm the $n=1$ photon ring emission, making the image-domain approach described here potentially infeasible for precise spin measurements.

% \clearpage
\vspace{1cm}

\section{Conclusions}
\label{sec:conclusions}

% motivation
The upcoming BHEX orbiting observation platform will enable an unprecedented view of the photon rings of the largest black holes on the sky. In particular, BHEX will probe the $n=1$ subring, which is largely invariant to surrounding astrophysical phenomenon and primarily affected by the black hole's spin and inclination. As a result, the $n=1$ photon ring has the potential to serve as a proxy for precisely measuring the spin of the black hole. However, the behavior of the $n=1$ subring as spin is varied is not completely understood, presenting a potential obstacle for tools hoping to exploit it to measure angular momentum. 

We present an end-to-end pipeline for characterizing the variation of black hole subrings as the properties of the black hole is modified. We take advantage of the fast semi-analytic simulation engine \texttt{KerrBAM} to generate a large volume ($\gtrsim10^6$) of $n=1$ subrings, varying spin and inclination in small increments. We develop a novel feature extraction algorithm (recursive brightest-point extraction) that is capable of extracting the non-circular brightness profiles of the subrings from these images on a short timescale, representing a $
\sim10^3$x performance increase over existing EHT-based tools. Next, we construct a series of geometric measurables which characterize the shape, orientation, and brightness distribution of the emission profile, and construct a large training set from these quantities.     

Our large dataset of $n=1$ subrings, combined with the minimal set of descriptive observables, provides an opportunity to presisely characterize the behavior of the $n=1$ subring. We quantify the behavior by fitting a bivariate cubic polynomial to each quantity and provide a mapping between combinations of $(a, \theta_0)$ and each geometric measurable. We then develop and apply a gradient boosted regression tree (GBRT) to the dataset of geometric measurables to broadly test whether these measurables are sufficient to constrain spin. We additionally test the performance of the GBRT against a convolutional neural network (CNN), which theoretically represents the maximum performance achievable on the images in our dataset (as the image necessarily contains more data than any set of geometric measurables). We find that, under no convolution, the GBRT and CNN are both able to measure spin from images with a high accuracy ($<10\%$ deviation from truth spin for all spins and inclinations). However, when convolution up to the BHEX beam ($\approx10 \mu$as) is introduced, both algorithms suffer at low spin and particularly low inclination, where the photon rings appear highly degenerate. Overall, the GBRT performs remarkably similarly to the CNN, indicating that the geometric measurables we have chosen successfully capture the salient features of the images. 

To more fully test the ability of the spin extraction methods described above, we test recovery on GRMHD simulations of black holes. These simulations are typically more accurate to reality but also more computationally expensive to generate. We chose simulations at the \virgoa inclination of $\theta_0=17^\circ$, which was particularly challenging for our algorithms during validation on the \texttt{KerrBAM} data. We begin by testing the GBRT and CNN algorithms on GRMHD simulations of the $n=1$ emission, to probe the utility of training on faster but potentially less accurate semi-analytic models. Remarkably, we find recovery of spin from the GRMHD images to be comparable in accuracy to the \texttt{KerrBAM} counterpart tests. Both the CNN and GBRT perform better at high spin and over-estimate the truth values at low spin.

We have presented a comprehensive pipeline for the study and analysis of black hole photon subrings. Analysis of the $n=1$ subring has illuminated aspects of how its geometry is modified with spin and inclination across the Kerr parameter space. Additionally, we have developed and trained several tools for extracting spin from semi-analytic and GRMHD simulations. Future research can expand upon any of these approaches in order to maximize utility to BHEX. The machine learning algorithms would benefit from an increased training set over a wider range of parameter combinations (e.g., electron heating) to both build more robust spin estimates and estimate parameters beyond just spin and inclination. To ensure that these procedures are effective, it is essential to adapt our methods to the specific observational constraints of BHEX. These tools are the basis of viable procedures to measure spin from future, sparse BHEX measurements of target black holes.

\acknowledgements{We thank the BHEX collaboration for useful discussions and feedback. JF is supported by an NSF-GRFP 2139319. J.D. is supported by a NASA Hubble Fellowship Program Einstein Fellowship. J.D. acknowledges support from a Joint Columbia University and Flatiron Institute Postdoctoral Fellowship. We thank Yaxuan Shen for helpful discussions regarding regularized gradient boosting.}

\bibliography{ref}

%% APPENDICES %%
% \clearpage
% \appendix
% \input{appendix}

\end{document}